\documentclass[]{JHEP3}
\usepackage{graphicx}
\usepackage{amsmath,amssymb,amscd,amsfonts,bm}

\newcommand{\as}{\alpha_{\mathrm{s}}}

\newcommand{\LA}{\mathrm{A}}

\newcommand{\LF}{\mathrm{F}}
\newcommand{\LI}{\mathrm{I}}
\newcommand{\LR}{\mathrm{R}}

\newcommand{\La}{\mathrm{a}}

\newcommand{\Lf}{\mathrm{f}}
\newcommand{\Lg}{\mathrm{g}}
\newcommand{\Lh}{\mathrm{h}}
\newcommand{\Lp}{\mathrm{p}}
\newcommand{\Ls}{\mathrm{s}}

\def\ket#1{\big|{#1}\big\rangle}
\def\bra#1{\big\langle{#1}\big|}
\def\brax#1{\big\langle{#1}}   

\def\sket#1{\big|{#1}\big)}
\def\sbra#1{\big({#1}\big|}
\def\sbrax#1{\big({#1}}        

\title{Final state dipole showers and the DGLAP equation}

\author{Zolt\'an Nagy \\
DESY\\
Notkestrasse 85\\
22607 Hamburg, Germany\\
E-mail: \email{Zoltan.Nagy@desy.de}
}

\author{Davison E. Soper\\
Institute of Theoretical Science\\
University of Oregon\\
Eugene, OR  97403-5203, USA\\
E-mail: \email{soper@uoregon.edu}
}

\abstract{
We study a parton shower description, based on a dipole picture, of the final state in electron-positron annihilation. In such a shower, the distribution function describing the inclusive probability to find a quark with a given energy depends on the shower evolution time. Starting from the exclusive evolution equation for the shower, we derive an equation for the evolution of the inclusive quark energy distribution in the limit of strong ordering in shower evolution time of the successive parton splittings. We find that, as expected, this is the DGLAP equation. This paper is a response to a recent paper of Dokshitzer and Marchesini that raised troubling issues about whether a dipole based shower could give the DGLAP equation for the quark energy distribution.
}

\preprint{DESY 08-208}

\begin{document}


\section{Introduction}

Parton shower algorithms, when coupled to hadronization models, provide a way of generating cross sections for exclusive final states according to approximations based on QCD. One can use these algorithms also to produce predictions for inclusive observables. Of special interest are predictions that, in a perturbative expansion, involve large logarithms of ratios of different momentum scales in the physical problem. In many cases, there are also predictions based on the full field theory. In these cases, it is of interest to know if a particular parton shower algorithm gives results consistent with the full field theory.

In this paper, we investigate one such case. We consider the inclusive parton energy distribution in electron-positron annihilation: the probability that a parton carries a fraction $x$ of the maximum possible energy of a parton. (We define the energy distribution precisely in Sec.~\ref{eq:Edistdef}.) We ask how the parton energy distribution changes as the resolution with which we look at the shower changes.  We use the shower evolution equation of Ref.~\cite{NSI}, which may be said to describe a dipole shower in that the emission of a gluon from a parton $l$ can interfere with the emission of a gluon from another parton $k$. The shower evolution equation describes the evolution of exclusive parton states. We manipulate the equation to produce an evolution equation for the inclusive parton energy distribution. This energy distribution of the partons should obey, approximately, the lowest order DGLAP evolution equation \cite{DGLAP} (like parton distribution functions).\footnote{The application to parton decay functions can be found in Ref.~\cite{FinalStateDGLAP}.}

This investigation is prompted by a recent paper of Dokshitzer and Marchesini \cite{Dokshitzer}, which raises some troubling issues with respect to the flow of momentum in the parton splittings. Ref.~\cite{Dokshitzer} suggests that perhaps a dipole shower cannot get the evolution of the energy distribution right. This is a significant issue since parton shower Monte Carlo event generators are of great importance in the design and interpretation of experiments and because some of these are based on a dipole picture. For instance, the program \textsc{Ariadne} \cite{Ariadne} is based on dipoles and, in fact, introduced the dipole idea as an organizing principle for parton showers.\footnote{However, we do not investigate whether the DGLAP equation results in \textsc{Ariadne}.} In particular, the recent program of Dinsdale, Ternick and Weinzierl \cite{Weinzierl} and the program of Schumann and Krauss \cite{Schumann} are both of the dipole type. These programs are based on the Catani-Seymour dipole splitting functions and momentum mapping \cite{CataniSeymour} that are often used in next-to-leading order perturbative calculations and can be applied to showers as described in Ref.~\cite{Ringberg}. Furthermore, the final state shower in version 8 of \textsc{Pythia} \cite{SjostrandSkands, Pythia8} is effectively a Catani-Seymour dipole shower. Thus if dipole showers cannot work for as significant a quantity as the evolution of the parton energy distribution, we ought to know about it.

For our investigation, we consider the process $e^+ + e^- \to {\it partons}$ in a shower picture. Our principle example is the shower evolution equation of Ref.~\cite{NSI}. The organizing principle of the shower is that one starts at the hard interaction $\gamma \to q + \bar q$ and moves to softer interactions, always factoring the softer interaction from previous harder interactions. We find in Sec.~\ref{sec:evolution} and Appendix \ref{sec:Eingluons} that with the definitions of Ref.~\cite{NSI}, we do obtain the expected DGLAP equations. We will also show in Sec.~\ref{CataniSeymour} that, at least for the case of $q \to q + \Lg$ splittings, the DGLAP equation results if one uses the splitting functions and momentum mapping of the Catani-Seymour dipole scheme. For either choice of splitting functions and momentum mapping, we take virtuality as the measure of hardness. One could also use the transverse momentum of the splittings to order successive splittings as in Refs.~\cite{Weinzierl, Schumann, SjostrandSkands, Pythia8}. As we outline in Secs. \ref{sec:PTordered} and \ref{CataniSeymour}, transverse momentum ordering also produces the DGLAP equation. We devote Sec.~\ref{sec:soft} to the role of soft gluons in the evolution equation, since this was the point of interest in Ref.~\cite{Dokshitzer}. 

The main point of this paper is to show that one can obtain the lowest order DGLAP equation from the shower evolution equation of Ref.~\cite{NSI} in a simple fashion, using the essential approximation that the shower splittings, which are by definition ordered in virtuality, are strongly ordered. In fact, successive splittings {\em are} strongly ordered if the splitting probability for each step $d q^2$ in virtuality is small, which follows if $\alpha_\Ls$ is small. 

\FIGURE{
\centerline{\includegraphics[width = 11 cm]{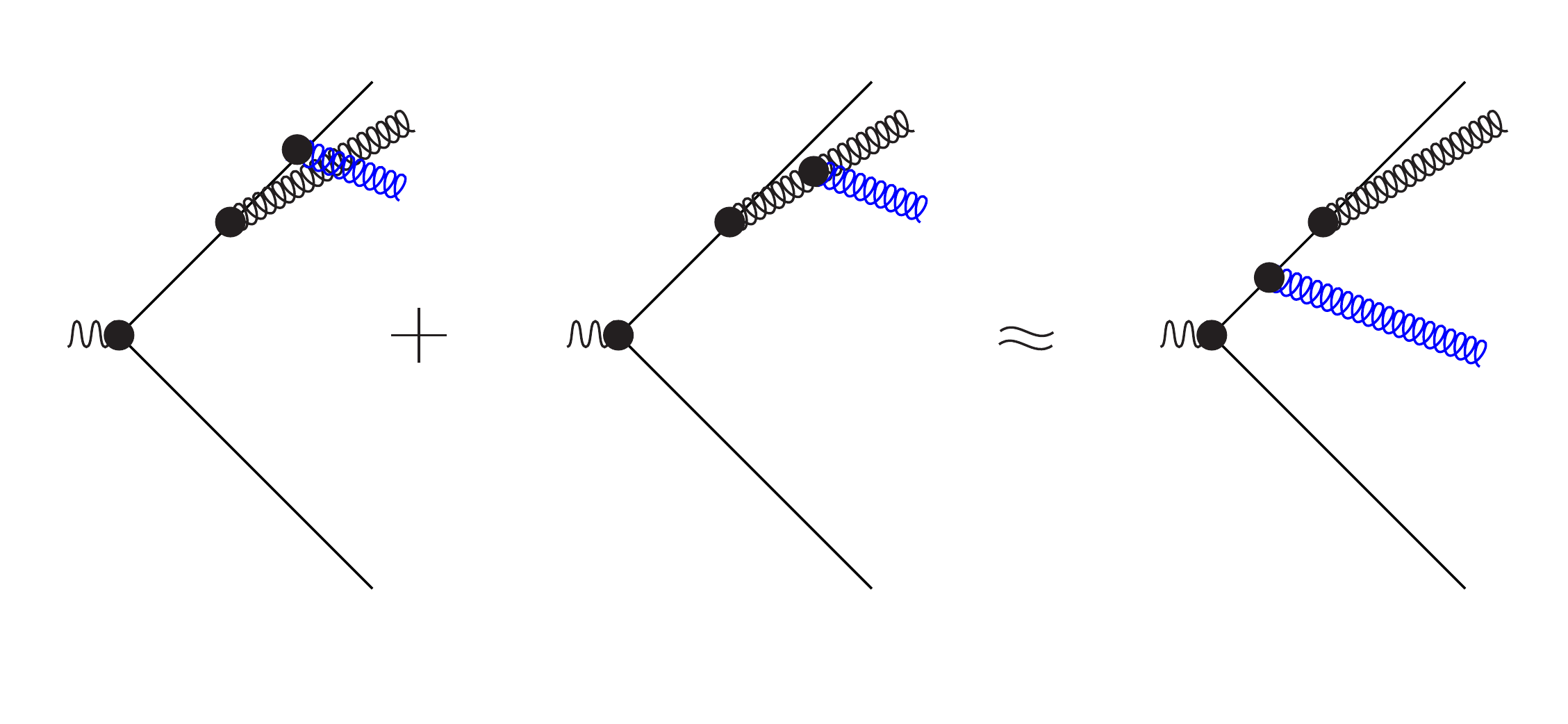}}
\caption{Emission of a soft gluon from a nearly collinear quark-gluon jet. The gluon is at a wider angle to the jet than the jet opening angle, but is very soft. The gluon does not resolve the jet, so the emission is as if the soft gluon were emitted from an on-shell massless parton moving in the direction of the jet and carrying the net color of the jet, as indicated on the right hand side of the approximation.}
\label{fig:softgluon1}
}

Before beginning the main analysis, we provide an informal and qualitative description of what we think the issue raised in Ref.~\cite{Dokshitzer} is. Suppose that in the first step of a shower description for $e^+ + e^- \to {\it partons}$, the initial quark emits a gluon with an energy $\omega_1$ equal to 1/3 of the initial quark energy, $E_q$. Then the daughter quark has energy $2/3\times E_q$. Suppose that the emission angle $\theta_1$ is very small, say $\theta_1 = 0.01$. Now suppose that a soft gluon with energy $\omega_2$ is emitted from the daughter quark or the daughter gluon at an angle $\theta_2$ that is large compared to $\theta_1$, say $\theta_2 = 0.1$. This is the second emission in a virtuality ordered shower, so the virtuality of the emission, $q_2^2 \sim E_q \omega_2 \theta_2^2$ must be smaller than the previous virtuality, $q_1^2 = E_q \omega_1 \theta_1^2$. Thus we need $\omega_2 < \omega_1 \theta_1^2/\theta_2^2 \sim 10^{-2} \omega_1$. 

\FIGURE{
\centerline{\includegraphics[width = 10 cm]{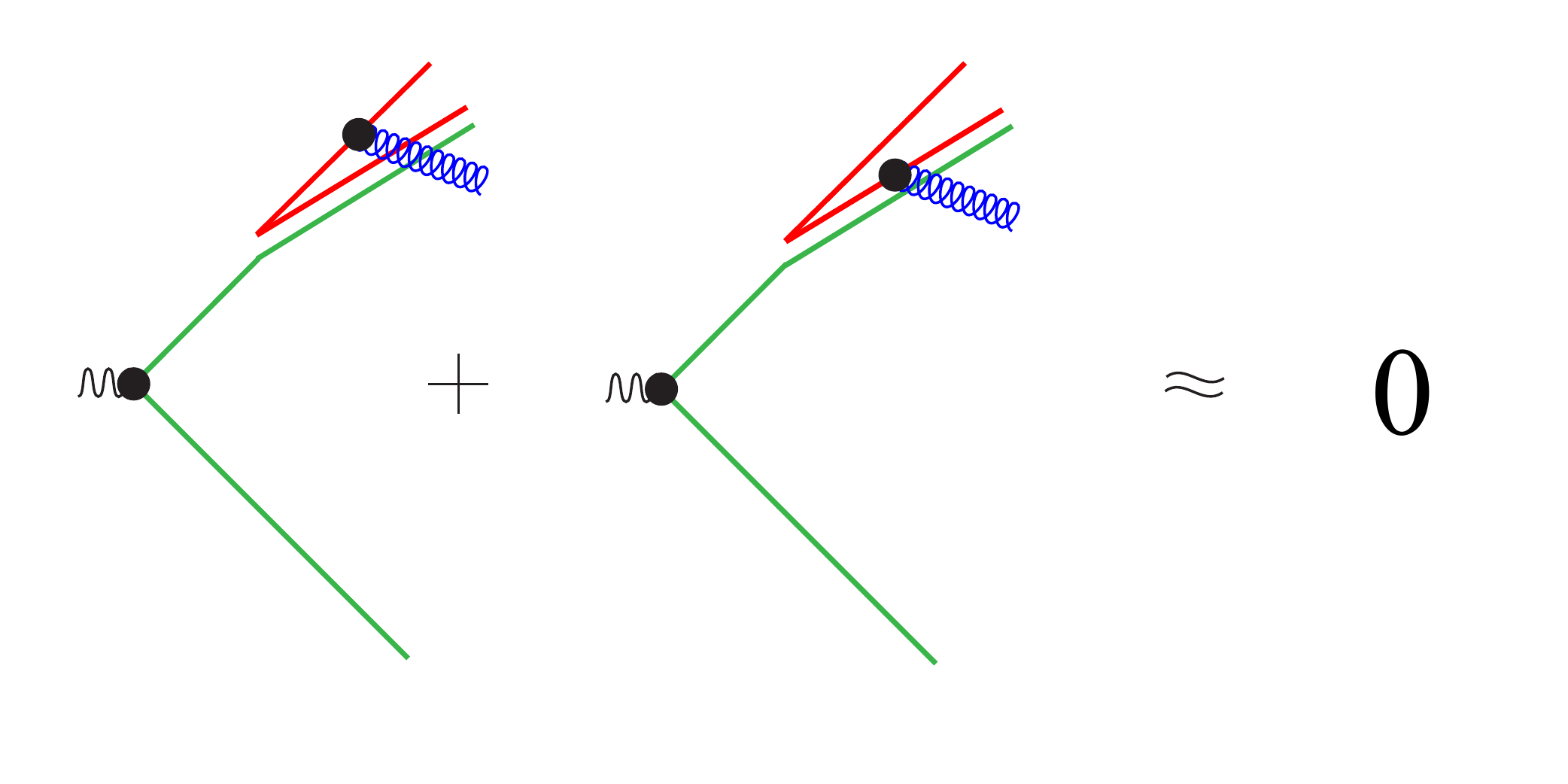}}
\caption{Emission of a soft gluon from a nearly collinear quark-gluon jet when the gluon in the jet is considered as ${\bm 3}\otimes \overline {\bm 3}$ state. Emission from the narrow dipole cancels.}
\label{fig:softgluon2}
}

Now we can ask where the energy of the soft gluon comes from. We note that the soft gluon cannot resolve the narrow jet. Thus, as indicated in Fig.~\ref{fig:softgluon1}, the matrix element for emitting the soft gluon is approximately the same as the matrix element for emitting the soft gluon from an on-shell quark moving in the direction of the mother quark. In an angular ordered shower, the emission is, in fact, modeled in this way \cite{angleorder}. This suggests that the energy of the soft gluon should come from the mother quark, so that 2/3 of it is taken from the daughter quark and 1/3 from the daughter gluon. However, this suggestion is (at least within the intuitive picture used here) misleading. If we make a simple leading color approximation, then the daughter gluon carries color  ${\bm 3}\otimes \overline {\bm 3}$ instead of color ${\bm 8}$. The soft gluon could be emitted from the $\overline {\bm 3}$ color in the daughter gluon, but the soft gluon cannot resolve this color from the ${\bm 3}$ color of the daughter quark. Thus this part of the gluon emission approximately cancels, as depicted in Fig.~\ref{fig:softgluon2}. Not all of the soft gluon emission amplitude cancels. The amplitude for emitting the soft gluon from the ${\bm 3}$ part of the daughter gluon, as depicted in Fig.~\ref{fig:softgluon3}, remains. This amplitude is approximately the same as that for emitting the soft gluon from the mother quark, as in the right hand side of Fig.~\ref{fig:softgluon1}. However, the soft gluon emission comes after the mother quark has split into a daughter quark and a daughter gluon. We thus see that it is the daughter gluon that gives up part of its energy to supply the energy of the soft gluon.

Does this issue matter? Perhaps, but it does not matter to the evolution of the energy distribution of the quark. The reason is that the energy taken by the soft, wide angle gluon is too tiny to make a difference in the limit in which we can regard the virtualities in the shower as strongly ordered. We will see this point appear in the actual shower evolution equations in the following sections.

\FIGURE{
\centerline{\includegraphics[width = 5 cm]{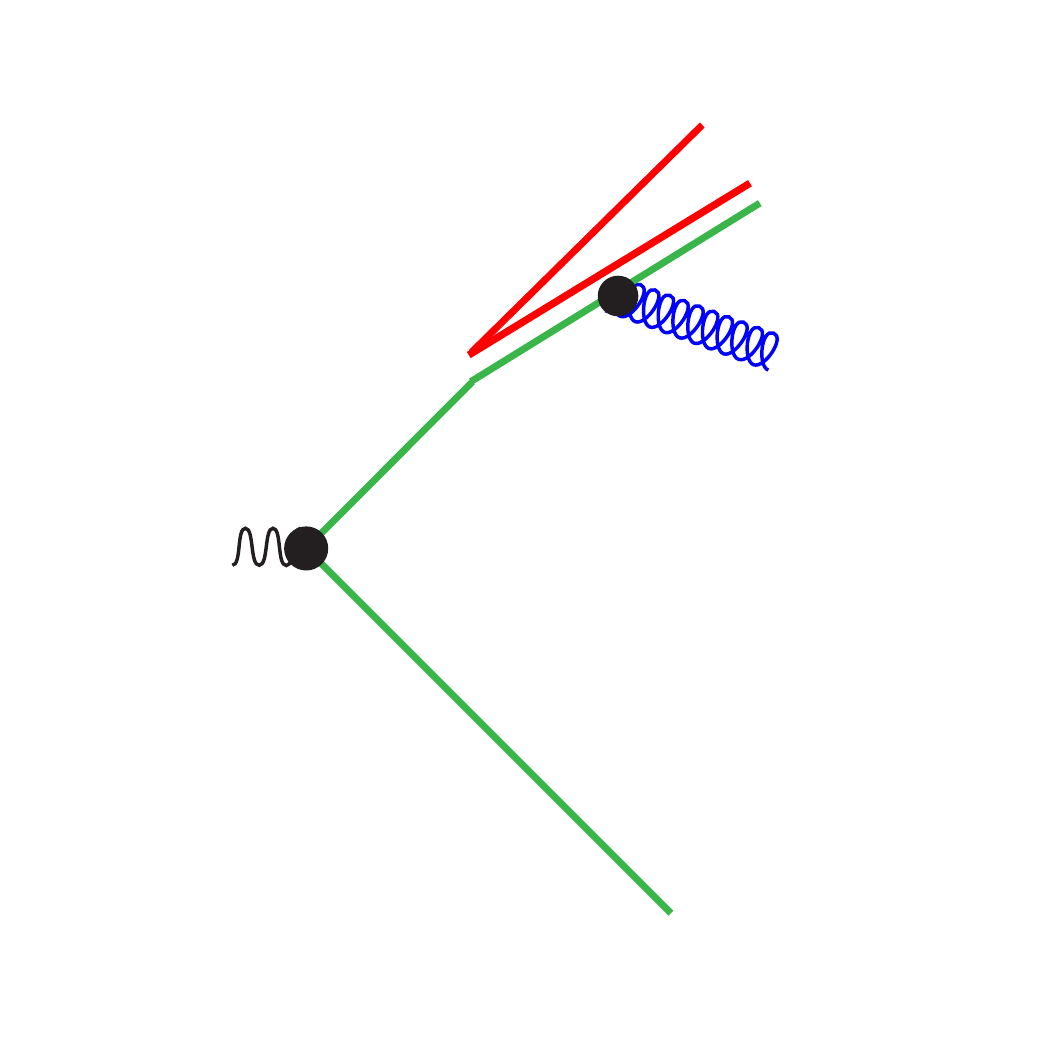}}
\caption{Emission of a soft gluon from a nearly collinear quark-gluon jet when the gluon in the get is considered as ${\bm 3}\otimes \overline {\bm 3}$ state. Emission from the part of the gluon that makes part of a wide angle dipole remains.}
\label{fig:softgluon3}
}

We point out that the virtuality ordering matters to this argument. Consider the case that $\omega_2 < \omega_1$ but $\omega_2 > \omega_1 \theta_1^2/\theta_2^2$. Then the soft gluon emission is harder than the collinear $q \to q + g$ splitting, so that the soft gluon emission belongs first in the shower, and part of its energy is to be deducted from what the daughter quark would otherwise get. Suppose that one were to use an energy ordered shower, despite the fact that the approximations used in factoring the softest splitting from harder splittings require that softer splittings are later in the shower. Then in a dipole style shower this soft gluon emission  would come second and the non-cancelling contribution would come from Fig.~\ref{fig:softgluon3}. The wrong parton would then be giving up its energy to supply the soft gluon energy. One would then not be surprised to not get the DGLAP evolution equation from such a shower evolution if one did not neglect $\omega_2$ compared to $E_q$. We suspect that something like this may have happened in Ref.~\cite{Dokshitzer}. In this paper, we use a virtuality ordered shower, as in Ref.~\cite{NSI}.

\section{The parton energy distribution}
\label{eq:Edistdef}

We consider the process $e^+ + e^- \to {\it partons}$ in a shower picture, using the notation of Ref.~\cite{NSI}. The shower evolution equation of Ref.~\cite{NSI} includes quantum interference among colors and spins. We note that this shower evolution equation is not directly practical for generating events. In order to implement the shower evolution equation, one would need an approximation scheme such that the approximated shower evolution can be practically implemented and such that the approximations can be systematically improved at a cost of computer time. In the lowest order approximation, one averages over spins and takes the leading color approximation, which yields an evolution equation that can be written as a Markov process \cite{NSII}. We will discover in this paper that we get the conventional DGLAP evolution equation for the energy distribution of partons as a function of resolution either with the full treatment of color and spin or in the leading color, spin averaged approximation.

In the notation of Ref.~\cite{NSI}, states in the sense of statistical mechanics are represented by ket vectors $\sket{\rho}$, while possible measurements are represented by bra vectors $\sbra{F}$. Thus $\sbrax{F}\sket{\rho}$ is the probability that one obtains a particular result $F$ from a measurement on an ensemble of systems represented by $\sket{\rho}$. We use basis states labelled by lists $\{p,f,s',c',s,c\}_{m}$ of parton quantum numbers. Here $\{p_1,\dots,p_m\} \equiv \{p\}_m$ are the momenta of the partons and $\{f\}_m$ are their flavors. The total momentum is $Q = \sum p_i$. The spins are specified by $\{s',s\}_{m}$, with two spin labels for each parton because we use the quantum density matrix in spin space in order to represent possible interference among spin states. The colors are similarly specified by $\{c',c\}_{m}$. Quantum spin and color states are represented by bra and ket vectors with angle brackets, as in the quantum spin inner product $\brax{\{s'\}_{m}}\ket{\{s\}_{m}}$. A much more detailed specification of our notation is provided in Ref.~\cite{NSI}. In Ref.~\cite{NSI}, the quarks can have non-zero masses, but in this paper we take all of the partons to be massless.

Let $\sket{\rho(t)} = {\cal U}(t,0)\sket{\rho(0)}$ be the statistical state at shower time $t$. Here $t=0$ at the beginning of the shower, which starts with the hard process, and $t \to \infty$ gives the statistical state after hadronization. The total cross section is
\begin{equation}
\begin{split}
\sbrax{1}\sket{\rho(t)}
={}&
\sum_m \frac{1}{m!}\int \big[d\{p,f,s',c',s,c\}_{m}\big]
\\&\quad\times
  \sbrax{1}\sket{\{p,f,s',c',s,c\}_{m}}
  \sbrax{\{p,f,s',c',s,c\}_{m}}\sket{\rho(t)}
\\={}&
\sum_m \frac{1}{m!}\int \big[d\{p,f,s',c',s,c\}_{m}\big]\,
  \brax{\{s'\}_{m}}\ket{\{s\}_{m}}\,
  \brax{\{c'\}_{m}}\ket{\{c\}_{m}}
\\ &\times
  \sbrax{\{p,f,s',c',s,c\}_{m}}\sket{\rho(t)}
\;\;.
\end{split}
\end{equation}
This is independent of $t$ because $\sbra{1}{\cal U}(t,0) = \sbra{1}$.
We now introduce a measurement function $\sbra{x,f_\La}$ that measures the probability to find in the final state a parton (or a hadron in the case $t \to \infty$) with flavor $f_\La$ having a fraction $x$ of the maximum possible energy for a single particle, $\sqrt {Q^2}/2$. The definition is
\begin{equation}
\begin{split}
\label{eq:xfrhodef}
\sbrax{x,f_\La}\sket{\rho(t)}
={}&
\sum_m \frac{1}{m!}\int \big[d\{p,f,s',c',s,c\}_{m}\big]
\\ &\quad\times
  \sbrax{x,f_\La}\sket{\{p,f,s',c',s,c\}_{m}}
  \sbrax{\{p,f,s',c',s,c\}_{m}}\sket{\rho}
\\={}&
\sum_m \frac{1}{m!}\int \big[d\{p,f,s',c',s,c\}_{m}\big]
  \brax{\{s'\}_{m}}\ket{\{s\}_{m}}\,
  \brax{\{c'\}_{m}}\ket{\{c\}_{m}}
\\ &\quad\times
\sum_{i=1}^m \delta_{f_\La,f_i}\,\delta\left(x - \frac{2p_i\cdot Q}{Q^2}\right)
  \sbrax{\{p,f,s',c',s,c\}_{m}}\sket{\rho(t)}
\;\;.
\end{split}
\end{equation}
The distribution measures all partons, so there is a sum over parton labels $i$. The energy fraction $x$ for parton $i$ is defined to be ${2p_i\cdot Q}/{Q^2}$.  Note that if we integrate over $x$ with a factor $x$ and sum over flavors, we get twice the total cross section: 
\begin{equation}
\begin{split}
\sum_{f_\La}\int_0^1\!dx\,
x\,\sbrax{x,f_\La}\sket{\rho(t)}
={}&
\sum_m \frac{1}{m!}\int \big[d\{p,f,s',c',s,c\}_{m}\big]\,
  \brax{\{s'\}_{m}}\ket{\{s\}_{m}}\,
  \brax{\{c'\}_{m}}\ket{\{c\}_{m}}
\\ &\times
  \left(\sum_i\frac{2 p_i\cdot Q}{Q^2}\right)\sbrax{\{p,f,s',c',s,c\}_{m}}\sket{\rho}
  \\ & = 2\, \sbrax{1}\sket{\rho(t)}
\;\;.
\end{split}
\end{equation}

When we apply this at $t = \infty$, we have a physical observable giving, say, the energy distribution of hadrons of flavor $\hat f_\Lh$ in an electron-positron collision, $\sbrax{\hat x,\hat f_\Lh}\sket{\rho(\infty)}$. This observable can be written in terms of the cross section to produce a parton of flavor $f_\La$ with a momentum fraction $x$ at a resolution scale $t$ convoluted with the parton decay function that gives the probability for that parton to decay to a hadron of the desired flavor carrying a fraction $\hat x/x$ of the parton's momentum:
\begin{equation}
\sbrax{\hat x,\hat f_\Lh}\sket{\rho(\infty)}
\approx \sum_{f_\La}\int_{\hat x}^1\frac{dx}{x}\ D_{\hat f_\Lh/f_\La}
\!\left(\frac{\hat x}{x},t\right)\,
\sbrax{x,f_\La}\sket{\rho(t)}
\;\;.
\end{equation}
This equation is approximate because factoring out the parton decay function involves neglecting power suppressed contributions. The parton decay functions depend on the resolution scale $t$ that we used to factor the cross section. This dependence is given by the DGLAP equation \cite{FinalStateDGLAP}. Since the observable $\sbrax{\hat x,\hat f_\Lh}\sket{\rho(\infty)}$ does not depend on $t$, the partonic function $\sbrax{x,f_\La}\sket{\rho(t)}$ must also depend on $t$ and this dependence must be given approximately by the DGLAP equation. We will investigate this evolution scale dependence within our dipole-like equations for shower evolution.

\section{Shower evolution for the energy distribution}
\label{sec:evolution}

We seek to discover how $\sbrax{\hat x,\hat f_\La}\sket{\rho(t)}$ evolves with shower resolution parameter $t$. We will consider the case in which $\hat f_\La$ is a quark flavor, $\hat f_\La \in {\cal Q} = \{u,d,s,\dots\}$. This was the case examined in Ref.~\cite{Dokshitzer}. The case of an antiquark is essentially identical. The case of a gluon is analyzed in Appendix \ref{sec:Eingluons}. 

From the evolution equation \cite{NSI} for $\sket{\rho(t)}$, we have 
\begin{equation}
\label{eq:xfevolution0}
\frac{d}{dt}\sbrax{\hat x,\hat f_\La}\sket{\rho(t)}
=
\sbra{\hat x,\hat f_\La}
{\cal H}_\LI(t) - {\cal V}(t)
\sket{\rho(t)}
\;\;.
\end{equation}
Here ${\cal H}_\LI(t)$ is the parton splitting operator and ${\cal V}(t)$ is the virtual ``no-splitting'' operator that insures that the probability for a splitting plus the probability for no splitting equals unity. We want to investigate $\sbra{\hat x,\hat f_\La} {\cal H}_\LI(t) - {\cal V}(t) \sket{\rho(t)}$ in the limit of strongly ordered virtualities in the shower. That is, the virtuality in the shower splitting described by ${\cal H}_\LI(t) - {\cal V}(t)$ should be small compared to the virtualities in the state $\sket{\rho(t)}$ before the splitting. We will make approximations that reflect this condition.

When we insert a complete sum over statistical states next to $\sket{\rho(t)}$ in Eq.~(\ref{eq:xfevolution0}), we get
\begin{equation}
\begin{split}
\label{eq:xfevolution1}
\frac{d}{dt}\sbrax{\hat x,\hat f_\La}\sket{\rho(t)}
={}&
\sum_m \frac{1}{m!}\int \big[d\{p,f,s',c',s,c\}_{m}\big]
\\ &\times
\sbra{\hat x,\hat f_\La}
{\cal H}_\LI(t) - {\cal V}(t)
\sket{\{p,f,s',c',s,c\}_{m}}
\\ & \times
\sbrax{\{p,f,s',c',s,c\}_{m}}
\sket{\rho(t)}
\;\;.
\end{split}
\end{equation}
In the following subsections, we will evaluate the matrix element of ${\cal H}_\LI(t) - {\cal V}(t)$ in the approximation of strongly ordered virtualities. In the last subsection, we will insert the limiting result for this matrix element back into Eq.~(\ref{eq:xfevolution1}). This will yield the DGLAP equation.

In the analysis of this section, we will freely make approximations that follow from having strongly ordered virtualities. Some of these approximations, related to soft, wide angle gluon emissions, were the main focus of Ref.~\cite{Dokshitzer}. In Sec.~\ref{sec:soft}, we turn to a detailed examination of these soft emission approximations. 

\subsection{The splitting matrix element}

To see what Eq.~(\ref{eq:xfevolution1}) means, we examine the matrix element of ${\cal H}_\LI(t) - {\cal V}(t)$ using the results in Sec.~12 of Ref.~\cite{NSI},
\begin{equation}
\begin{split}
\label{eq:HVxf1}
\sbra{\hat x,\hat f_\La}&
{\cal H}_\LI(t)
-{\cal V}(t)
\sket{\{p,f,s',c',s,c\}_{m}}
=
\\&
\sum_l
\sum_{\zeta_{\rm f}\in \Phi_{l}(f_l)}
\int d\zeta_{\rm p}\
\theta(\zeta_{\rm p} \in \varGamma_{l}(\{p\}_{m},\zeta_{\rm f}))\,
\delta\!\left(
t - T_l(\{\hat p,\hat f\}_{m+1})
\right)
\\&\times 
\bigg[
\sum_{i=1}^{m+1}
\delta_{\hat f_\La,\hat f_i}\,
\delta\!\left(\hat x - \frac{2\hat p_i\cdot Q}{Q^2}\right)
-
\sum_{i=1}^m
\delta_{\hat f_\La,f_i}\,
\delta\!\left(\hat x - \frac{2 p_i\cdot Q}{Q^2}\right)
\bigg]
\\&\times
\biggl\{
\theta(\hat f_{m+1}\ne {\rm g})\
T_\LR
\brax{\{c'\}_{m}}
\ket{\{c\}_{m}}\
\brax{\{s'\}_{m}}
\ket{\{s\}_{m}}\
\overline w_{ll}(\{\hat f,\hat p\}_{m+1})
\\ & +
\theta(\hat f_{m+1} = {\rm g})
\sum_{k \ne l}
\bra{\{c'\}_{m}}
t_k(f_k \to \hat f_k + \hat f_{m+1})\,
t^\dagger_l(f_l \to \hat f_l + \hat f_{m+1})
\ket{\{c\}_{m}}
\\&\quad
\times 
\brax{\{s'\}_{m}}\ket{\{s\}_{m}}\
\bigg[2
A_{lk}(\{\hat p\}_{m+1})
\overline w_{lk}(\{\hat f,\hat p\}_{m+1})
-
\overline w_{ll}(\{\hat f,\hat p\}_{m+1})
\bigg]
\bigg\}
\;\;.
\end{split}
\end{equation}
This formula is a little complicated because it contains several ingredients. However, its important features are easily understood.

In the second line, there is a sum over the index $l$ of the parton that splits. The daughter partons are labelled $l$ and $m+1$. Then there is a sum over the flavors of the daughter partons into which parton $l$ splits. These flavors are denoted by $\zeta_\Lf = (\hat f_l, \hat f_{m+1})$. If $f_l$ is a quark or antiquark flavor, it must split into the same flavor quark or antiquark plus a gluon. If $f_l$ is a gluon, it can split into two gluons or a quark-antiquark pair. Then there is an integration over splitting variables that determine the momenta of the daughters. These can be a virtuality $y$, a momentum fraction $z$, and an azimuthal angle $\phi$. This integration is indicated as $d\zeta_\Lp$. The momenta $\hat p_i$ of the partons after the splitting are given in terms of the momenta $p_i$ before the splitting together with the splitting variables $\zeta_{\rm p}$. Finally in the second line there is a delta function that defines the shower time $t$, for which we will use the virtuality of the splitting,
\begin{equation}
\label{eq:Tdef}
T_l(\{\hat p,\hat f\}_{m+1}) = \log\left(
\frac{Q^2}{2\hat p_l\cdot \hat p_{m+1}}
\right)
\;\;.
\end{equation}

In the third line there is a difference of two terms. The first, with $\hat f_i$ and $\hat p_i$, comes from ${\cal H}_\LI(t)$. The second, with $f_i$ and $p_i$, comes from ${\cal V}(t)$. In this virtual contribution, the energy fraction $\hat x$ is set to the energy fraction of parton $i$ before the splitting and we integrate inclusively over the splitting variables.

The functions that follow are defined in Ref.~\cite{NSI}. They are derived from the spin dependent splitting functions for the splitting of a parton with label $l$ to produce two daughter partons with, in our labeling scheme, labels $l$ and $m+1$. 

The line starting $\theta(\hat f_{m+1}\ne {\rm g})$ is for a $\Lg \to q + \bar q$ splitting. It has a color factor $T_\LR \brax{\{c'\}_{m}}\ket{\{c\}_{m}}$ with $T_\LR = 1/2$. The function $\overline w_{ll}(\{\hat f,\hat p\}_{m+1})$ is the spin-averaged splitting function for $\Lg \to q + \bar q$, as given in Ref.~\cite{NSI}. This term is depicted in Fig.~\ref{fig:gqq}. The dependence on the mother parton spin is trivial, $\brax{\{s'\}_{m}}\ket{\{s\}_{m}}$, because we sum over the daughter parton spins and integrate over the azimuthal angle of the splitting, as indicated in Eq.~(12.18) of  Ref.~\cite{NSI}.

\FIGURE{
\centerline{\includegraphics[width = 10 cm]{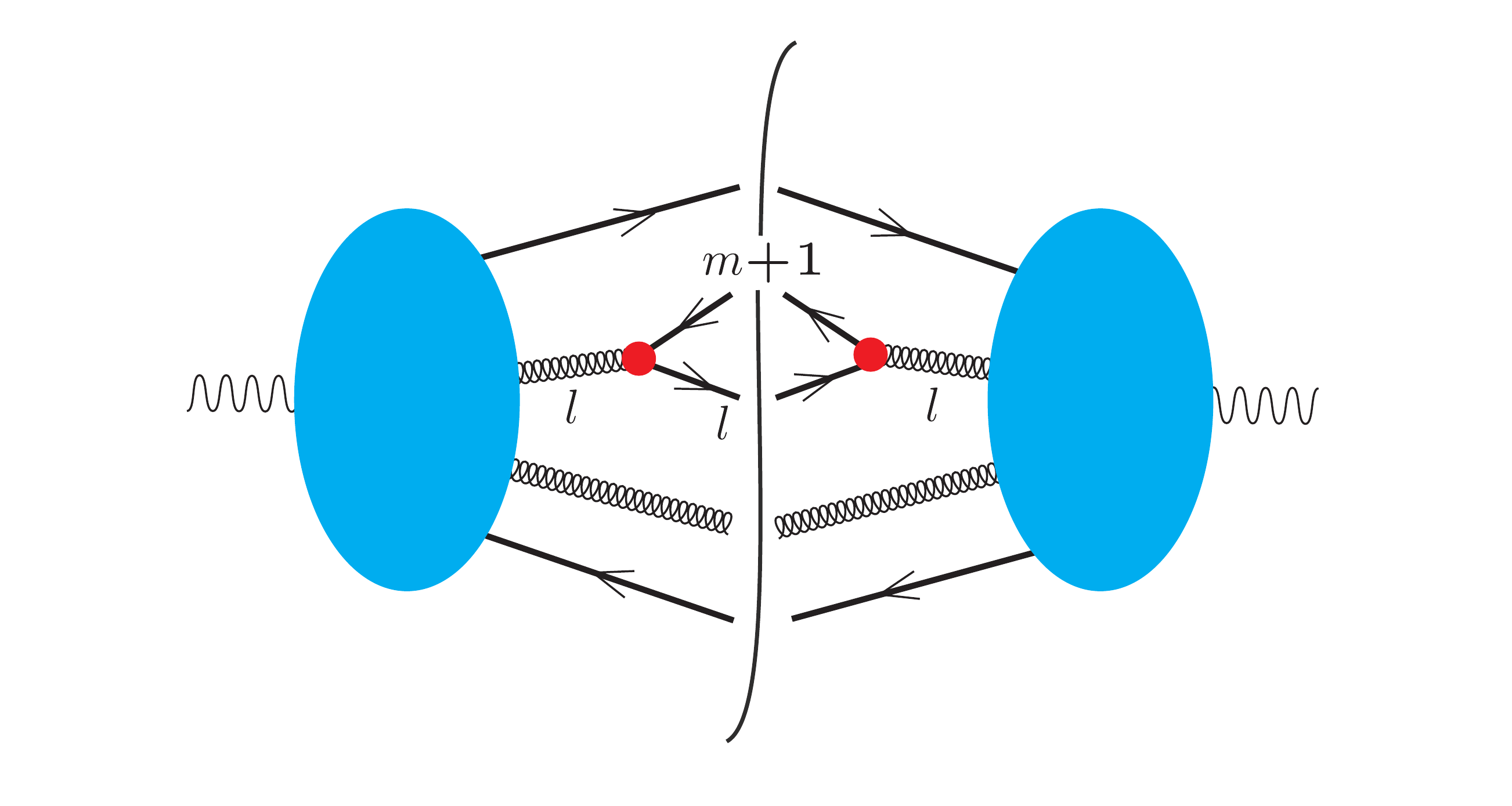}}
\caption{A graph contributing to $\overline w_{ll}$ for a $\Lg \to q + \bar q$ splitting.}
\label{fig:gqq}
}

\FIGURE{
\centerline{\includegraphics[width = 10 cm]{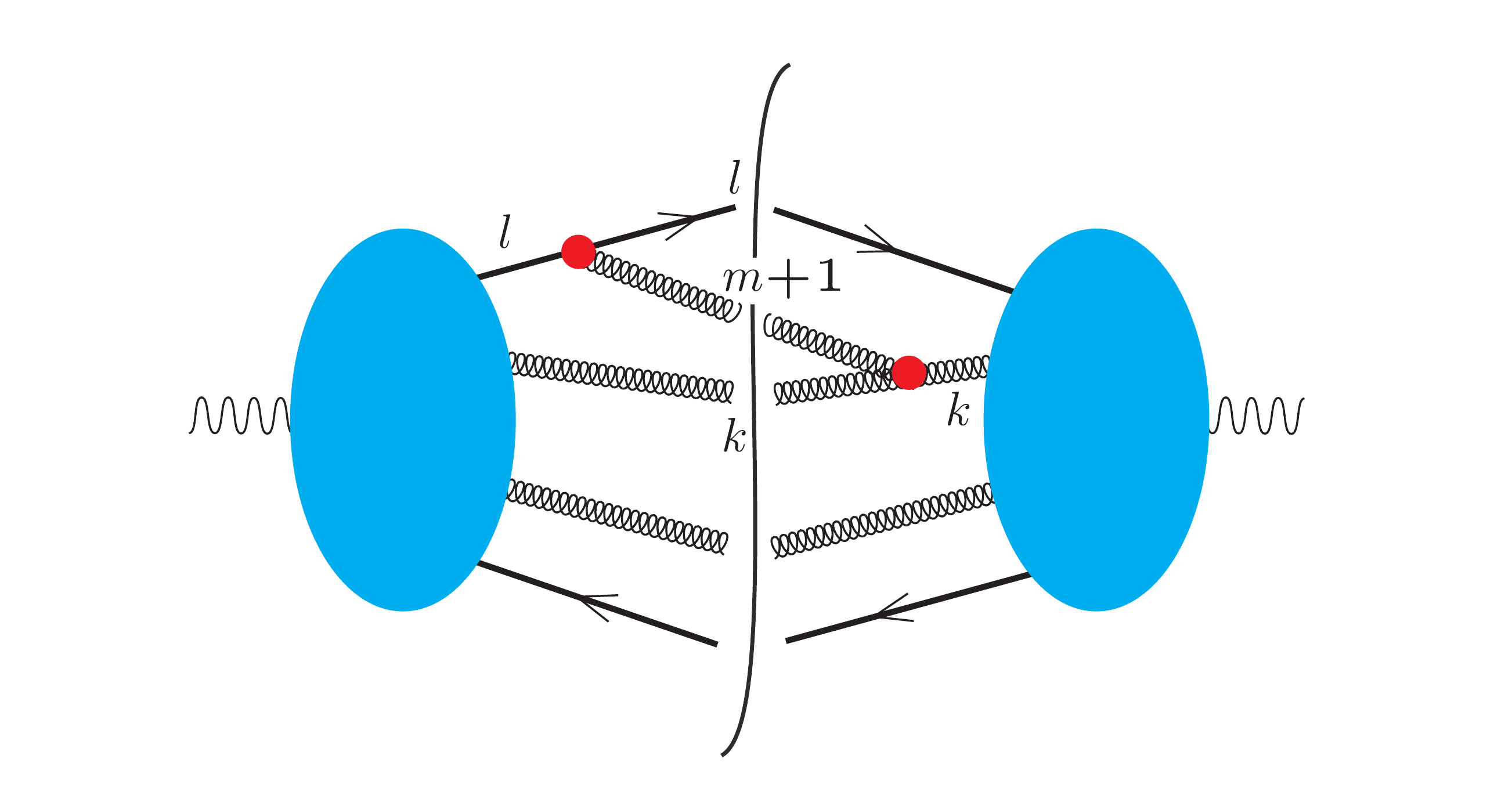}}
\caption{An interference graph contributing to $\overline w_{lk}$ for a $q \to q + \Lg$ splitting.}
\label{fig:qqginterference}
}

\FIGURE{
\centerline{\includegraphics[width = 10 cm]{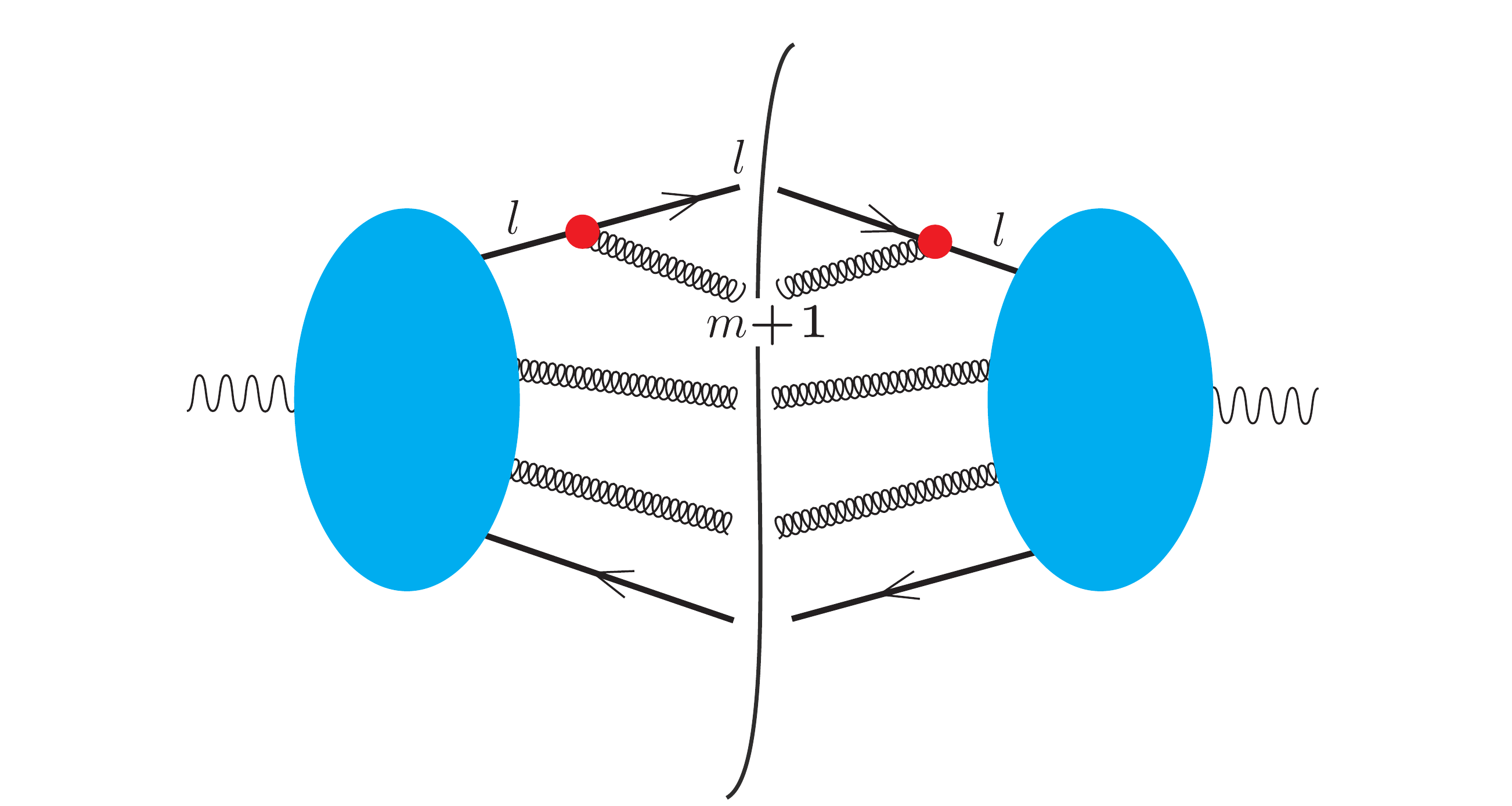}}
\caption{A direct graph contributing to $\overline w_{ll}$ for a $q \to q + \Lg$ splitting.}
\label{fig:qqgdirect}
}

The lines starting $\theta(\hat f_{m+1} = {\rm g})$ are for a $q \to q + \Lg$ splitting. It has a nontrivial color structure $\bra{\{c'\}_{m}}
t_k\, t^\dagger_l \ket{\{c\}_{m}}$. This corresponds to an interference graph and is the color factor for emitting a gluon from parton $l$ in the quantum amplitude and absorbing it on line $k$ in the complex conjugate amplitude.\footnote{This factor is written as the matrix element of $\bm{T}_k \cdot \bm{T}_l$ in the work of Catani and Seymour \cite{CataniSeymour}.} This color factor multiplies a sum of two terms. The first, containing $\overline w_{lk}$, corresponds to the interference graphs, depicted in Fig.~\ref{fig:qqginterference}. The second, containing $\overline w_{ll}$, corresponds to the direct graphs, depicted in Fig.~\ref{fig:qqgdirect}. The functions $\overline w_{lk}$ and $\overline w_{ll}$ are given in Ref.~\cite{NSI}, while function $A_{lk}$, or rather possible choices for it, are given in Ref.~\cite{NSIII}.

\subsection{Strongly ordered virtualities}

We can immediately simplify Eq.~(\ref{eq:xfevolution1}) in the strong ordering approximation. There are sums over partons $i$ that participate in the measurement represented by $\sbra{\hat x,\hat f_\La}$. Thus $\hat x$ is set equal to $2\hat p_i\cdot Q/Q^2$ in the term from ${\cal H}_\LI(t)$ and $2p_i\cdot Q/Q^2$ in the term from ${\cal V}(t)$. Consider the contribution if $i$ is the same in both terms and is neither of the indices of the daughter partons, $l$ or $m+1$. For a splitting that is nearly collinear or is very soft, or both, we have $\hat p_i \approx p_i$ for such a spectator parton. In fact, we would have $\hat p_i = p_i$ except for the need to take some momentum from the spectator partons in order to balance momentum in the splitting while keeping both the daughter partons and the mother parton on shell. The amount of momentum needed tends to zero as the virtuality in the splitting approaches zero. For this reason, $\hat p_i - p_i \to 0$ in the limit of small virtuality in the splitting. Thus in the limit of strongly ordered virtualities in the shower, we drop the terms for $i$ not equal to $l$ or $m+1$. This becomes yet simpler because we consider the case that $\hat f_\La$ is a quark flavor. With our labeling scheme, parton $m+1$ is always either a gluon or (for $\Lg \to q + \bar q$) an antiquark. Thus, in fact, the only term that remains in the sum over $i$ is that for $i=l$. This gives
\begin{equation}
\begin{split}
\label{eq:HVxf2}
\sbra{\hat x,\hat f_\La}&
{\cal H}_\LI(t)
-{\cal V}(t)
\sket{\{p,f,s',c',s,c\}_{m}}
\approx
\\&
\sum_l
\sum_{\zeta_{\rm f}\in \Phi_{l}(f_l)}
\int d\zeta_{\rm p}\
\theta(\zeta_{\rm p} \in \varGamma_{l}(\{p\}_{m},\zeta_{\rm f}))\,
\delta\!\left(
t - T_l(\{\hat p,\hat f\}_{m+1})
\right)\
\brax{\{s'\}_{m}}\ket{\{s\}_{m}}\
\\&\times 
\bigg[
\delta_{\hat f_\La,\hat f_l}\,
\delta\!\left(\hat x - \frac{2\hat p_l\cdot Q}{Q^2}\right)
-
\delta_{\hat f_\La,f_l}\,
\delta\!\left(\hat x - \frac{2 p_l\cdot Q}{Q^2}\right)
\bigg]
\\&\times
\biggl\{
\theta(\hat f_{m+1}\ne {\rm g})\
T_\LR
\brax{\{c'\}_{m}}
\ket{\{c\}_{m}}\
\overline w_{ll}(\{\hat f,\hat p\}_{m+1})
\\ & +
\theta(\hat f_{m+1} = {\rm g})
\sum_{k \ne l}
\bra{\{c'\}_{m}}
t_k(f_k \to \hat f_k + \hat f_{m+1})\,
t^\dagger_l(f_l \to \hat f_l + \hat f_{m+1})
\ket{\{c\}_{m}}
\\&\quad
\times 
\bigg[2
A_{lk}(\{\hat p\}_{m+1})
\overline w_{lk}(\{\hat f,\hat p\}_{m+1})
-
\overline w_{ll}(\{\hat f,\hat p\}_{m+1})
\bigg]
\bigg\}
\;\;.
\end{split}
\end{equation}

The bookkeeping for the flavor structure can now be made more explicit. Consider the two terms on the third line of Eq.~(\ref{eq:HVxf2}), representing a real splitting and a virtual splitting. If the mother parton $l$ is a gluon, the virtual splitting contribution vanishes because $\hat f_\La \ne f_l$. In the real splitting contribution, the Kroneker delta restricts the sum over flavors $\zeta_\Lf$ in the splitting to just one term, $(\hat f_l, \hat f_{m+1}) = (\hat f_\La, -\hat f_\La)$. If parton $l$ is not a gluon, it must in fact be a quark with flavor $f_l = \hat f_\La$. Then again, there is only one term in the sum over flavors $\zeta_\Lf$ in the splitting, namely $(\hat f_l, \hat f_{m+1}) = (\hat f_\La, \Lg)$. Thus
\begin{equation}
\begin{split}
\label{eq:HVxf3}
\sbra{\hat x,\hat f_\La}&
{\cal H}_\LI(t)
-{\cal V}(t)
\sket{\{p,f,s',c',s,c\}_{m}}
\approx
\\&
\sum_l
\int d\zeta_{\rm p}\
\theta(\zeta_{\rm p} \in \varGamma_{l}(\{p\}_{m},\zeta_{\rm f}))\,
\delta\!\left(
t - T_l(\{\hat p,\hat f\}_{m+1})
\right)
\sum_{\hat f_l,\hat f_{m+1}}
\brax{\{s'\}_{m}}\ket{\{s\}_{m}}
\\&\times
\biggl\{
\theta(f_{l} = {\rm g},\hat f_l = \hat f_\La,
       \hat f_{m+1} = - \hat f_\La)
\\&\qquad\times
T_\LR
\brax{\{c'\}_{m}}
\ket{\{c\}_{m}}\
\overline w_{ll}(\{\hat f,\hat p\}_{m+1})\
\delta\!\left(\hat x - \frac{2\hat p_l\cdot Q}{Q^2}\right)
\\ &\quad +
\theta(f_{l} = \hat f_\La,\hat f_l = \hat f_\La,
       \hat f_{m+1} = \Lg)
\\ &\qquad\times
\sum_{k \ne l}
\bra{\{c'\}_{m}}
t_k(f_k \to \hat f_k + \hat f_{m+1})\,
t^\dagger_l(f_l \to \hat f_l + \hat f_{m+1})
\ket{\{c\}_{m}}
\\&\qquad
\times 
\bigg[2
A_{lk}(\{\hat p\}_{m+1})
\overline w_{lk}(\{\hat f,\hat p\}_{m+1})
-
\overline w_{ll}(\{\hat f,\hat p\}_{m+1})
\bigg]
\\&\qquad\times
\bigg[
\delta\!\left(\hat x - \frac{2\hat p_l\cdot Q}{Q^2}\right)
-
\delta_{\hat f_\La,f_l}\,
\delta\!\left(\hat x - \frac{2 p_l\cdot Q}{Q^2}\right)
\bigg]
\bigg\}
\;\;.
\end{split}
\end{equation}

A further simplification is possible. The interference diagram of Fig.~\ref{fig:qqginterference}, represented by the function $\overline w_{lk}$, is important in the development of the exclusive final state in a parton shower. However, $\overline w_{lk}$ is singular only in the limit of a {\em soft} gluon emission. It is not singular for emission with $\hat p_{m+1}$ almost parallel to $p_l$. In the limit of a soft gluon emission, the soft gluon does not take any momentum from parton $l$. Thus $\hat p_l \approx p_l$, so that the two terms on the last line of Eq.~(\ref{eq:HVxf3}) cancel. Thus, for our inclusive observable, in the strong ordering limit the term in Eq.~(\ref{eq:HVxf3}) proportional to $\overline w_{lk}$ can be dropped. We recognize that this step touches on the main point of Ref.~\cite{Dokshitzer}, so we will return to provide a more detailed analysis of it in Sec.~\ref{sec:soft}.

Once we have dropped $\overline w_{lk}$, the sum over $k$ in  Eq.~(\ref{eq:HVxf3}) is trivial. As in Ref.~\cite{CataniSeymour} and in Eq.~(8.25) of Ref.~\cite{NSI}, we use the fact that the complete quantum state is a color singlet to write
\begin{equation}
\begin{split}
\sum_{k \ne l}
\bra{\{c'\}_{m}}&
t_k(f_k \to \hat f_k + \hat f_{m+1})\,
t^\dagger_l(f_l \to \hat f_l + \hat f_{m+1})
\ket{\{c\}_{m}}
\\
={}& - \bra{\{c'\}_{m}}
t_l(f_k \to \hat f_k + \hat f_{m+1})\,
t^\dagger_l(f_l \to \hat f_l + \hat f_{m+1})
\ket{\{c\}_{m}}
\\
={}& - C_\LF\,
\brax{\{c'\}_{m}}
\ket{\{c\}_{m}}
\;\;.
\end{split}
\end{equation}
This simplification holds when we treat color exactly. It also holds when we use the leading color approximation, as explained in Ref.~\cite{NSII}.

It will also prove useful to define $x = {2 p_l\cdot Q}/{Q^2}$ and introduce into Eq.~(\ref{eq:HVxf2}) a factor
\begin{equation}
1 = \int_0^1\! dx \ 
\delta\!\left(x - \frac{2 p_l\cdot Q}{Q^2}\right)
\;\;.
\end{equation}
With these changes, Eq.~(\ref{eq:HVxf3}) becomes
\begin{equation}
\begin{split}
\label{eq:HVxf4}
\sbra{\hat x,\hat f_\La}&
{\cal H}_\LI(t)
-{\cal V}(t)
\sket{\{p,f,s',c',s,c\}_{m}}
\approx
\\&
\sum_l
\int d\zeta_{\rm p}\
\theta(\zeta_{\rm p} \in \varGamma_{l}(\{p\}_{m},\zeta_{\rm f}))\,
\delta\!\left(
t - T_l(\{\hat p,\hat f\}_{m+1})
\right)
\sum_{\hat f_l,\hat f_{m+1}}
\\&\times
\brax{\{s'\}_{m}}\ket{\{s\}_{m}}\
\brax{\{c'\}_{m}}\ket{\{c\}_{m}}\
\int_0^1\! dx \ 
\delta\!\left(x - \frac{2 p_l\cdot Q}{Q^2}\right)
\\&\times
\biggl\{
\theta(f_{l} = {\rm g}, \hat f_l = \hat f_\La,
       \hat f_{m+1} = - \hat f_\La)\
T_\LR\,
\overline w_{ll}(\{\hat f,\hat p\}_{m+1})\,
\delta\!\left(\hat x - \frac{\hat p_l\cdot Q}{p_l\cdot Q}\,x\right)
\\ &\quad
+
\theta(f_{l} = \hat f_\La, \hat f_l = \hat f_\La,
      \hat f_{m+1} = \Lg)\
\\&\qquad\times
C_\LF\,
\overline w_{ll}(\{\hat f,\hat p\}_{m+1})
\bigg[
\delta\!\left(\hat x - \frac{\hat p_l\cdot Q}{p_l\cdot Q}\,x\right)
-
\delta\!\left(\hat x - x\right)
\bigg]
\bigg\}
\;\;.
\end{split}
\end{equation}

\subsection{Kinematics}

At this point, we introduce the variables used in Ref.~\cite{NSII} to describe the splitting of parton $l$ with momentum $p_l$ to produce daughter parton $l$ with momentum $\hat p_l$ and daughter parton $m+1$ with momentum $\hat p_{m+1}$. We define a virtuality variable
\begin{equation}
y = \frac{2 \hat p_l\cdot \hat p_{m+1}}{2 p_l\cdot Q}
\;\;.
\end{equation}
We use virtuality as the evolution variable, as defined in Eq.~(\ref{eq:Tdef}). (In Sec.~\ref{sec:PTordered}, we examine using transverse momentum as the evolution variable.) Accordingly, the delta function that sets the shower time of the splitting gives
\begin{equation}
\label{eq:ydef}
y = \frac{Q^2}{2 p_l\cdot Q}\ e^{-t}
\;\;.
\end{equation}
We define a lightlike vector $n_l$ in the $p_l$-$Q$ plane and use this vector to define a momentum fraction variable $z$ by
\begin{equation}
z = \frac{\hat p_l\cdot n_l}{\hat p_l\cdot n_l + \hat p_{m+1}\cdot n_l}
\;\;.
\end{equation}
These are the variables used in Ref.~\cite{NSII} except that we have interchanged $z \leftrightarrow (1-z)$ in order to follow the convention for a $q \to q + \Lg$ splitting that $z$ is the momentum fraction of the daughter quark. There is a third splitting variable, the azimuthal angle of the transverse part of $\hat p_{m+1}$ in a reference frame in which $p_l$ and $n_l$ define the 0-3 plane.

We can work out from the definition in Ref.~\cite{NSI} that if we use these splitting variables,
\begin{equation}
\int d\zeta_\Lp\,
\theta(\zeta_{\rm p} \in \varGamma_{l}(\{p\}_{m},\zeta_{\rm f}))
\cdots\ 
= \frac{p_l\cdot Q}{8\pi^2}\int_0^{y_{\rm max}}\!dy\ \lambda
\int_0^1\!dz\, \int \!\frac{d\phi}{2\pi}\cdots
\;\;.
\end{equation}
Here,\footnote{In this paper, we denote $2p_l\cdot Q/Q^2 = x$. In Ref.~\cite{NSI} we wrote $2p_l\cdot Q/Q^2 = 1/a_l$.} 
\begin{equation}
\begin{split}
\lambda ={}& \sqrt{(1+y)^2 - 4 y/x}
\;\;,
\\
y_{\rm max} ={}& \left(
\sqrt{\frac{1}{x}}
- \sqrt{\frac{1}{x} - 1}\,
\right)^2
\;\;.
\end{split}
\end{equation}
We take note that $x < 1$ so that $\lambda < 1$. The maximum value of $y$ is the value that makes $\lambda = 0$. Also $1-\lambda$ is proportional to $y$ for $y \to 0$. Using the delta function that sets $y$ in terms of $t$, we have
\begin{equation}
\begin{split}
\label{eq:zetapandt}
\int d\zeta_\Lp\,
\theta(\zeta_{\rm p} \in \varGamma_{l}(\{p\}_{m},\zeta_{\rm f}))\
&\delta\!\left(
t - T_l(\{\hat p,\hat f\}_{m+1})
\right)\cdots\ 
\\ ={}&
\frac{p_l\cdot Q}{8\pi^2}\ y\,\theta(y < y_{\rm max})\
\lambda
\int_0^1\!dz\, \int \!\frac{d\phi}{2\pi}\cdots
\;\;,
\end{split}
\end{equation}
where now $y$ is given by Eq.~(\ref{eq:ydef}). When we take the limit $y \to 0$, we have $\theta(y < y_{\rm max}) \to 1$. We also have $\lambda \to 1$.

The delta function $\delta\!\left(\hat x - ({\hat p_l\cdot Q}/{p_l\cdot Q})\,x\right)$ relates $\hat x = {2\hat p_l\cdot Q}/Q^2$ to $x = {2 p_l\cdot Q}/Q^2$ and the splitting variables $y$ and $z$ according to the momentum mapping in Ref.~\cite{NSI}. The relation is determined from
\begin{equation}
\label{eq:xtohatx}
\frac{\hat p_l\cdot Q}{p_l\cdot Q} = z + \frac{1}{2}\,y
-\left(z - \frac{1}{2}\right)\, (1 - \lambda)
\;\;.
\end{equation}
We will use the full relation in Sec.~\ref{sec:soft}. In this section, we immediately take the limit of strongly ordered virtualities, $y \to 0$,
\begin{equation}
\frac{\hat p_l\cdot Q}{p_l\cdot Q} \approx z
\;\;.
\end{equation}
Thus
\begin{equation}
\begin{split}
\label{eq:HVxf5}
\sbra{\hat x,\hat f_\La}&
{\cal H}_\LI(t)
-{\cal V}(t)
\sket{\{p,f,s',c',s,c\}_{m}}
\approx
\\&
\sum_l
\frac{p_l\cdot Q}{8\pi^2}\ y
\int_0^1\!dz\, \int \!\frac{d\phi}{2\pi}
\sum_{\hat f_l,\hat f_{m+1}}
\\&\times
\brax{\{s'\}_{m}}\ket{\{s\}_{m}}\
\brax{\{c'\}_{m}}\ket{\{c\}_{m}}\
\int_0^1\! dx \ 
\delta\!\left(x - \frac{2 p_l\cdot Q}{Q^2}\right)
\\&\times
\biggl\{
\theta(f_{l} = {\rm g}, \hat f_l = \hat f_\La,
       \hat f_{m+1} = - \hat f_\La)\
T_\LR\,
\overline w_{ll}(\{\hat f,\hat p\}_{m+1})\,
\frac{1}{z}\,
\delta\!\left(x - \frac{\hat x}{z}\right)
\\ &\quad
+
\theta(f_{l} = \hat f_\La, \hat f_l = \hat f_\La,
      \hat f_{m+1} = \Lg)\
\\&\qquad\times
C_\LF\,
\overline w_{ll}(\{\hat f,\hat p\}_{m+1})
\bigg[
\frac{1}{z}\,
\delta\!\left(x - \frac{\hat x}{z}\right)
-
\delta\!\left(x - \hat x\right)
\bigg]
\bigg\}
\;\;.
\end{split}
\end{equation}

\subsection{Splitting functions}

We can take the functions $\overline w_{ll}$ from Ref.~\cite{NSII}. They have the form of a simple prefactor times a function of the splitting variables $y$ and $z$ and the mother parton energy fraction $x$,
\begin{equation}
\label{eq:wll}
\overline w_{ll}(\{\hat f,\hat p\}_{m+1})
= \frac{4\pi\alpha_\Ls}{p_l\cdot Q\ y}\times
\begin{cases} 
F_{q/g}(y,z,x), &
(\hat f_l, \hat f_{m+1}) \in \{({\rm u},\bar {\rm u}), ({\rm d},\bar {\rm d}), \dots\} \\
F_{q/q}(y,z,x), &
(\hat f_l, \hat f_{m+1}) \in \{({\rm u}, \Lg), ({\rm d},\Lg), \dots\} 
\end{cases}
\;\;.
\end{equation}
The functions $F$ are, from Eqs.~(A.1) and (2.23) of Ref.~\cite{NSII},
\begin{equation}
\begin{split}
\label{eq:FqgFqq}
F_{q/g}(y,z,x) ={}& 
z^2 + (1-z)^2
\;\;,
\\
F_{q/q}(y,z,x) ={}& 
\frac{(1 + \lambda + y)^2}{4\lambda}\ 
\left[
\frac{2\,\tilde z}{1-\tilde z}
- \frac{2 y}{x(1-\tilde z)^2 (1+y)^2}
\right]
\\&
+\frac{1}{2} (1-z)[1+y + \lambda]
\;\;,
\end{split}
\end{equation}
where\footnote{The variable $\tilde z$ is called $x$ in Ref.~\cite{NSII}, Eq.~(2.21). However, in this paper, we use $x$ for a different purpose.}
\begin{equation}
\label{eq:tildezdef}
\tilde z(y,z,x) = \frac{\lambda z + \frac{1}{2}(1 + y - \lambda)}{1+y}
\;\;.
\end{equation}
Note that $\tilde z(0,z,x) = z$. The function $F_{q/g}$ is independent of $y$ and $x$, while $F_{q/q}$ has a simple, $x$ independent, limit as $y \to 0$. We have
\begin{equation}
\begin{split}
\label{eq:splittingFlimits}
F_{q/g}(0,z,x) ={}& 
z^2 + (1-z)^2
\;\;,
\\
F_{q/q}(0,z,x) ={}& \frac{1 + z^2}{1-z}
\;\;.
\end{split}
\end{equation}
Note that these limiting forms are the DGLAP kernels. In fact, one might say that the limit of the splitting functions as $y \to 0$ at fixed $z$ {\em defines} the DGLAP kernels for final state evolution.

Using this result, performing the trivial $\phi$-integration, and taking the $y \to 0$ limit, we have
\begin{equation}
\begin{split}
\label{eq:HVxf6}
\sbra{\hat x,\hat f_\La}&
{\cal H}_\LI(t)
-{\cal V}(t)
\sket{\{p,f,s',c',s,c\}_{m}}
\approx
\\&
\sum_l
\frac{\alpha_\Ls}{2\pi}
\int_0^1\!dz\,
\int_0^1\!dx \ 
\delta\!\left(x - \frac{2 p_l\cdot Q}{Q^2}\right)
\brax{\{s'\}_{m}}\ket{\{s\}_{m}}\
\brax{\{c'\}_{m}}\ket{\{c\}_{m}}\
\\&\times
\biggl\{
\theta(f_{l} = {\rm g})\
T_\LR\,
[z^2 + (1-z)^2]\
\frac{1}{z}\,
\delta\!\left(x - \frac{\hat x}{z}\right)
\\&\quad
+
\theta(f_{l} = \hat f_\La)\
C_\LF\,
\frac{1 + z^2}{1-z}\,
\bigg[
\frac{1}{z}\,
\delta\!\left(x - \frac{\hat x}{z}\right)
-
\delta\!\left(x - \hat x\right)
\bigg]
\bigg\}
\;\;.
\end{split}
\end{equation}

\subsection{The DGLAP equation}

We insert Eq.~(\ref{eq:HVxf6}) into Eq.~(\ref{eq:xfevolution1}) and use the definition (\ref{eq:xfrhodef}) of $\sbrax{x, f}\sket{\rho(t)}$ on the right hand side. We obtain
\begin{equation}
\begin{split}
\label{eq:xfevolution2}
\frac{d}{dt}\sbrax{\hat x,\hat f_\La}\sket{\rho(t)}
\approx{}&
\frac{\alpha_\Ls}{2\pi}\int_0^1\!dz\
\bigg\{
T_\LR\,
[z^2 + (1-z)^2]\,
\frac{1}{z}\,\sbrax{\hat x /z, \Lg}\sket{\rho(t)}
\\&\quad +
C_\LF\,
\frac{1 + z^2}{1-z}
\Big[
\frac{1}{z}\,\sbrax{\hat x /z, \hat f_\La}\sket{\rho(t)}
-\sbrax{\hat x, \hat f_\La}\sket{\rho(t)}
\Big]
\bigg\}
\;\;.
\end{split}
\end{equation}
This is typically written
\begin{equation}
\begin{split}
\label{eq:xfevolution3}
\frac{d}{dt}\sbrax{\hat x,\hat f_\La}\sket{\rho(t)}
\approx{}&
\frac{\alpha_\Ls}{2\pi}\int_{\hat x}^1\!\frac{dz}{z}\
\bigg\{
T_\LR\,
[z^2 + (1-z)^2]\,
\sbrax{\hat x /z, \Lg}\sket{\rho(t)}
\\&\quad +
C_\LF\,
\left[\frac{1 + z^2}{1-z}\right]_+
\sbrax{\hat x /z, \hat f_\La}\sket{\rho(t)}
\bigg\}
\;\;.
\end{split}
\end{equation}
This is the DGLAP equation for $\sbrax{\hat x,\hat f_\La}\sket{\rho(t)}$ where $\hat f_\La$ is a quark flavor. It has one term for $\Lg \to q + \bar q$ splitting and one for $q \to q + \Lg$ splitting. The ``+'' prescription corresponds to the subtraction Eq.~(\ref{eq:xfevolution2}). It comes from the virtual splitting operator ${\cal V}(t)$ in Eq.~(\ref{eq:xfevolution0}).

\subsection{Transverse momentum ordered shower}
\label{sec:PTordered}

We have chosen virtuality as the shower ordering variable. One could use transverse momentum instead,
\begin{equation}
\label{eq:TdefPT}
T_l(\{\hat p,\hat f\}_{m+1}) = \log\left(
\frac{Q^2}{2z(1-z)\hat p_l\cdot \hat p_{m+1}}
\right)
\;\;.
\end{equation}
In this case, it is
\begin{equation}
y_T = z (1-z) y
= \frac{Q^2}{2 p_l\cdot Q} \ e^{-t}
\end{equation}
that is fixed by the delta function $\delta(t - T_l)$. The strong ordering limit is then $y_T \to 0$ at fixed $z$. In this case, one should include a factor
\begin{equation}
\theta(z(1-z) > y_T/y_{\rm max})
\end{equation}
that limits the $z$ integral in both the real and virtual splitting terms. However, this factor disappears in the limit $y_T \to 0$, leaving us with the same result (\ref{eq:xfevolution2}).

\section{Soft gluon emission}
\label{sec:soft}

In this section, we examine the effects of soft, wide angle gluon emission. In Sec.~\ref{sec:evolution}, we argued that soft gluon emission could be ignored. Thus we dropped the interference term containing $\overline w_{lk}$ and, for the direct term $\overline w_{ll}$ for a $q \to q + \Lg$ splitting, we took the limit $y \to 0$ with fixed $z$. The argument was that for a soft splitting, the observed quark feels no recoil, so that the virtual splitting term cancels the real splitting term. However, Ref.~\cite{Dokshitzer} argues that a dipole shower could fail to reproduce the DGLAP equation when the quark recoil is accounted for. Accordingly, we examine the contribution from soft, wide angle gluon emission when we account for the recoil using the momentum mapping from Ref.~\cite{NSI}. We write the $q \to q + \Lg$ contribution exactly, then examine the contribution in the soft emission limit. We find that this contribution vanishes  in the limit of strongly ordered virtualities, $y \to 0$.

We begin with Eq.~(\ref{eq:HVxf3}) for the matrix element of ${\cal H}_\LI(t)
-{\cal V}(t)$. There is a term for a $\Lg \to q + \bar q$ splitting and a term for a $q \to q + \Lg$ splitting. We examine the $q \to q + \Lg$ term. We use the definitions for $d\zeta_\Lp$ and the evolution variable $t$, Eq.~(\ref{eq:zetapandt}), as before. We introduce an integral of a delta function that defines the variable $x$ as before. We also introduce an integral of a delta function that defines the angle $\vartheta$ between $p_l$ and $p_k$. This gives
\begin{equation}
\begin{split}
\label{eq:qqg1}
\sbra{\hat x,\hat f_\La}&
{\cal H}_\LI(t)
-{\cal V}(t)
\sket{\{p,f,s',c',s,c\}_{m}}_{q \to q + \Lg}
=
\\&
\sum_l\sum_{k \ne l}
\frac{p_l\cdot Q}{8\pi^2}\ y\,\theta(y < y_{\rm max})\ \lambda
\int_0^1\!dz\, \int \!\frac{d\phi}{2\pi}\
\sum_{\hat f_l,\hat f_{m+1}}
\brax{\{s'\}_{m}}\ket{\{s\}_{m}}
\\ & \times
\int_0^1\!dx\
\delta\!\left(x - \frac{2 p_l\cdot Q}{Q^2}\right)\,
\int_{-1}^1 d\cos\vartheta\
\delta\!\left(
\cos\vartheta
-1
+\frac{p_l\cdot p_k\, Q^2}{p_l\cdot Q\ p_k\cdot Q}
\right)
\\&\times
\theta(f_{l} = \hat f_\La,\hat f_l = \hat f_\La, \hat f_{m+1} = \Lg)
\\&\times
\bra{\{c'\}_{m}}
t_k(f_k \to \hat f_k + \hat f_{m+1})\,
t^\dagger_l(f_l \to \hat f_l + \hat f_{m+1})
\ket{\{c\}_{m}}
\\&
\times 
\bigg[2
A_{lk}(\{\hat p\}_{m+1})
\overline w_{lk}(\{\hat f,\hat p\}_{m+1})
-
\overline w_{ll}(\{\hat f,\hat p\}_{m+1})
\bigg]
\\&\times 
\bigg[
\delta\!\left(\hat x - \frac{\hat p_l\cdot Q}{p_l\cdot Q}\,x\right)
-
\delta\!\left(\hat x - x\right)
\bigg]
\;\;.
\end{split}
\end{equation}

\subsection{The splitting functions}

For $\overline w_{ll}$, we use the exact expression given by Eqs.~(\ref{eq:wll}) and (\ref{eq:FqgFqq}). For $\overline w_{lk}$, we define
\begin{equation}
\label{eq:wlk}
2\,A_{lk}\overline w_{lk} = 
\frac{4\pi\alpha_\Ls}{p_l\cdot Q\ y}\ 
F_{{\rm int}}(y,z,\phi;x,\vartheta)
\;\;.
\end{equation}
The $q \to q + \Lg$ interference function $F_{{\rm int}}$ includes the function $A_{lk}$ that describes how we divide the $l$-$k$ interference term between a contribution that we associate with parton $l$ and another that we associate with parton $k$. There is some freedom in choosing $A_{lk}$. For our purposes here, we choose the version specified in Eqs.~(7.2) and (7.12) of Ref.~\cite{NSIII}. (Any of the other choices based on Eq.~(7.2) from Ref.~\cite{NSIII} would do as well.) The function $F_{{\rm int}}$ is simple when written in terms of the momenta $\hat p_l$, $\hat p_{m+1}$, $\hat p_k$ and $Q$. When expressed as a function of $x,\vartheta$ and the splitting variables $y,z,\phi$ according to the momentum mapping of Ref.~\cite{NSI}, $F_{{\rm int}}$ is a little complicated. We find
\begin{equation}
\begin{split}
\label{eq:Fint}
F_{{\rm int}}(&y,z,\phi; x,\vartheta) =
\\&
2-
\frac{2[g_1(y,1-z,x,\vartheta) + g_1(y,z,x,\vartheta)]\,
g_3(y,z) + 2y\,g_2(y,x,\vartheta)}
{g_1(y,z,x,\vartheta)g_3(y,z)  + y\, g_2(y,x,\vartheta)
- 2\,\tan(\vartheta/2) \sqrt{x z(1-z)y}\,g_3(y,z) \, \cos \phi}
\\&
+\frac{2\tilde z}{1-\tilde z}
- \frac{2y}{x(1-\tilde z)^2 (1+y)^2}
\;\;,
\end{split}
\end{equation}
where $\tilde z$ was defined in Eq.~(\ref{eq:tildezdef}), and where
\begin{equation}
\begin{split}
\label{eq:g123}
g_1(y,z,x,\vartheta) ={}&  
\frac{z\, x}{2( 1 - x)}\,[1 - y - \lambda]
+ \frac{1}{2}\,x \tan^2(\vartheta/2)\,(1-z)\,[1-y+\lambda]
\;\;,
\\
g_2(y,x,\vartheta) ={}&  
 \frac{2 - (1 + y)x - \lambda\,x}{2(1 - x)} 
+ \frac{1}{2}\, \tan^2(\vartheta/2)\,[2 -(1 + y)x +\lambda\,x]
\;\;,
\\
g_3(y,z) ={}&  
\frac{z}{2}\,[1+y+\lambda] 
+\frac{1-z}{2}\,[1+y-\lambda]
\;\;.
\end{split}
\end{equation}

A simple calculation from Eqs.~(\ref{eq:Fint}) and (\ref{eq:g123}) shows that 
$F_{{\rm int}}$ is simple in the collinear limit, $y \to 0$ with $z$ fixed:
\begin{equation}
\label{eq:Fintcollinear}
F_{{\rm int}}(0,z,\phi; x,\vartheta) = 0
\;\;.
\end{equation}
This, of course, was obvious from the beginning. In a physical gauge like the Coulomb gauge used for our splitting functions, the interference graphs do not give a leading collinear singularity. Because of Eq.~(\ref{eq:Fintcollinear}), we neglected the $\overline w_{lk}$ term in the previous section. In this section, we do not simply take the limit $y \to 0$ limit with $z$ fixed.

\subsection{Kinematics}

The delta function $\delta\!\left(\hat x - ({\hat p_l\cdot Q}/{p_l\cdot Q})\,x\right)$ in Eq.~(\ref{eq:qqg1}) relates $\hat x = {2\hat p_l\cdot Q}/Q^2$ to $x = {2 p_l\cdot Q}/Q^2$ and the splitting variables $y$ and $z$ according to Eq.~(\ref{eq:xtohatx}). We can solve this relation for $x$, yielding
\begin{equation}
x = X(y,z,\hat x)
\;\;.
\end{equation}
The exact relation is rather complicated,
\begin{equation}
\begin{split}
X(y,z,\hat x) ={}& \frac{1}{2 z(1-z)(1+y)^2}\bigg\{
 (1+y)\hat x - y(2z-1)^2
\\ & 
-(2z-1) \sqrt{
\big[(1+y)\hat x - y\big]^2 - 4 y^2 z(1-z)
}
\bigg\}
\;\;.
\end{split}
\end{equation}
As $y \to 0$, $X(y,z,\hat x)$ approaches a very simple result
$X(0,z,\hat x) = {\hat x}/{z}$. We rewrite the delta function as
\begin{equation}
\label{eq:delta}
\delta\!\left(\hat x - ({\hat p_l\cdot Q}/{p_l\cdot Q})\,x\right)
= 
\frac{\partial X(y,z,\hat x)}{\partial \hat x}\,
\delta\!\left(x - X(y,z,\hat x)\right)
\;\;.
\end{equation}

\subsection{Evolution equation}

Using Eq.~(\ref{eq:delta}) and the substitutions for $\overline w_{lk}$ and $\overline w_{ll}$, Eq.~(\ref{eq:qqg1}) becomes
\begin{equation}
\begin{split}
\label{eq:qqg2}
\sbra{\hat x,\hat f_\La}&
{\cal H}_\LI(t)
-{\cal V}(t)
\sket{\{p,f,s',c',s,c\}_{m}}_{q \to q + \Lg}
=
\\&
\sum_l\sum_{k \ne l}
\frac{\alpha_\Ls}{2\pi}\,\theta(y < y_{\rm max})\ \lambda
\int_0^1\!dz\, \int \!\frac{d\phi}{2\pi}\
\brax{\{s'\}_{m}}\ket{\{s\}_{m}}
\\ & \times
\int_0^1\!dx\
\delta\!\left(x - \frac{2 p_l\cdot Q}{Q^2}\right)\,
\int_{-1}^1 d\cos\vartheta\
\delta\!\left(
\cos\vartheta
-1
+\frac{p_l\cdot p_k\, Q^2}{p_l\cdot Q\ p_k\cdot Q}
\right)
\\&\times
\theta(f_{l} = \hat f_\La)
\bra{\{c'\}_{m}}
t_k(f_k \to f_k + \Lg)\,
t^\dagger_l(f_l \to  f_l + \Lg)
\ket{\{c\}_{m}}
\\&
\times 
\bigg[F_{{\rm int}}(y,z,\phi;x,\vartheta)
-
F_{q/q}(y,z,x)
\bigg]
\\&\times 
\bigg[
\frac{\partial X(y,z,\hat x)}{\partial \hat x}\,
\delta\!\left(x - X(y,z,\hat x)\right)
-
\delta\!\left(\hat x - x\right)
\bigg]
\;\;.
\end{split}
\end{equation}
We insert this into Eq.~(\ref{eq:xfevolution1}), giving the $q\to q + \Lg$ contribution to the evolution of the energy distribution function. This contribution then takes the form
\begin{equation}
\begin{split}
\label{eq:xfevolutionsoft1}
\left[
\frac{d}{dt}\sbrax{\hat x,\hat f_\La}\sket{\rho(t)}
\right]_{q\to q + \Lg}
={}&
\int_0^1\!dz\, \int \!\frac{d\phi}{2\pi}
\int_{-1}^1\! d\cos\vartheta\
\theta(y < y_{\rm max})
\\&\times
\frac{\alpha_\Ls}{2 \pi}\,\lambda
\bigg[
F_{{\rm int}}(y,z,\phi;x,\vartheta)
-
F_{q/q}(y,z,x)
\bigg]
\\&\times
\bigg[
\frac{\partial X(y,z,\hat x)}{\partial \hat x}\,
\sbrax{X(y,z,\hat x),\hat f_\La,\vartheta}\sket{\rho(t)}
\\ & \qquad
-
\sbrax{\hat x,\hat f_\La,\vartheta}\sket{\rho(t)}
\bigg]
\;\;,
\end{split}
\end{equation}
where $\sbrax{x, f_\La,\vartheta}\sket{\rho(t)}$ is a two parton correlation function,
\begin{equation}
\begin{split}
\label{eq:2partonfctndef}
\sbrax{x, f_\La,\vartheta}\sket{\rho(t)} ={}& 
\sum_m \frac{1}{m!}\int \big[d\{p,f,s',c',s,c\}_{m}\big]
\\ &\times
\sum_l \sum_{k \ne l}
\delta\!\left(x - \frac{2 p_l\cdot Q}{Q^2}\right)
\delta\!\left(
\cos\vartheta
-1
+\frac{Q^2\, p_l\cdot p_k}{Q\cdot p_l\ Q\cdot p_k}
\right)
\\&\times
\theta(f_{l} =  f_\La)\,
\brax{\{s'\}_{m}}\ket{\{s\}_{m}}
\\&\times
\bra{\{c'\}_{m}}
t_k(f_k \to f_k + \Lg)\,
t_l^\dagger(f_l \to f_l + \Lg)
\ket{\{c\}_{m}}
\\&\times
\sbrax{\{p,f,s',c',s,c\}_{m}}
\sket{\rho(t)}
\;\;.
\end{split}
\end{equation}
Thus we find an evolution equation similar to Eq.~(\ref{eq:xfevolution2}) except that the evolution of the one parton distribution is expressed in terms of a more complicated function, a two parton distribution.

\subsection{The soft emission limit}

If we were to follow the analysis of the Sec.~\ref{sec:evolution}, we would take the limit $y\to 0$ at fixed $z$ in the right hand side of Eq.~(\ref{eq:xfevolutionsoft1}). However, we should be a little careful about this. The kernel is singular not only for $y\to 0$ at fixed $z$ but also for $y \to 0$ with $(1-z) = \xi y$ with $\xi$ fixed. This is the limit of soft, wide angle gluon emission when expressed using the variables $y$, $z$. Define the gluon emission angle $\vartheta_{m+1}$ by
\begin{equation}
1-\cos \vartheta_{m+1} = \frac{\hat p_{m+1}\cdot p_l\ Q^2}
{\hat p_{m+1}\cdot Q\ p_l\cdot Q}
\;\;,
\end{equation}
Then $\xi$ is related to $\vartheta_{m+1}$ by
\begin{equation}
\xi \equiv \frac{1-z}{y}
= \left[
\frac{1}{2}\left[
1 + y^2 + (1+y)\lambda - \frac{2y}{x}
\right]
\tan^2(\vartheta_{m+1}/2) + \frac{2 y}{1 + \cos \vartheta_{m+1}}
\right]^{-1}
\;\;.
\end{equation}
The range of $\xi$ is $0 < \xi < 1/y$, with $\xi \to 0$ when $\vartheta_{m+1} \to \pi$ and $\xi \to 1/y$ when $\vartheta_{m+1} \to 0$. For $y \to 0$ at fixed $\vartheta_{m+1}$,
\begin{equation}
\xi \sim \frac{1}{\tan^2(\vartheta_{m+1}/2)}
\;\;.
\end{equation}
Our purpose in this section is to investigate whether the soft emission integration region, $y \to 0$ at finite $\xi$, contributes to the evolution at small $y$.

The integration over the soft region can be expressed by using $\xi$ as the integration variable, with
\begin{equation}
dz = y\,d\xi
\;\;.
\end{equation}
Taking $y \to 0$ at finite $\xi$, the factor $y$ from the integration measure times the kernel becomes
\begin{equation}
\begin{split}
\label{eq:softlimit}
y\,\lambda
\bigg[
F_{{\rm int}}(y,z,\phi;x,\vartheta)
&-
F_{q/q}(y,z,x)
\bigg]
\\
\sim{}&
-\frac{2 x \tan^2(\vartheta/2)}
{2 + (1 + x \xi)\tan^2(\vartheta/2) 
- 2 \tan(\vartheta/2) \sqrt{x\xi} \cos \phi}
\;\;.
\end{split}
\end{equation}
Thus we would have a finite contribution to the integration from the soft emission region in the limit $y \to 0$ if we were to multiply the splitting function by anything that is finite in the soft emission limit.

However, for our inclusive observable, we need to multiply by the function
\begin{equation}
\label{eq:subtractedfactor}
\left[
\frac{\partial X(y,z,\hat x)}{\partial \hat x}\
\sbrax{X(y,z,\hat x),\hat f_\La, \vartheta}\sket{\rho(t)}
-
\sbrax{\hat x,\hat f_\La, \vartheta}\sket{\rho(t)}
\right]
\;\;.
\end{equation}
Using the expansion
\begin{equation}
X(y,z,\hat x) \sim \hat x 
+ y\,(1 - \hat x + \xi \hat x)
+ {\cal O}(y^2)
\;\;,
\end{equation}
we see that the factor (\ref{eq:subtractedfactor}) becomes in the soft limit
\begin{equation}
y\,\frac{\partial}{\partial \hat x}
\left[
(1 - \hat x + \xi \hat x)
\sbrax{\hat x,\hat f_\La, \vartheta}\sket{\rho(t)}
\right]
+ {\cal O}(y^2)
\;\;.
\end{equation}
Thus the integrand has a net factor $y$ in the soft limit, so that the soft region does not contribute to the leading shower evolution of $\sbrax{\hat x,\hat f_\La}\sket{\rho(t)}$. (We have investigated the $q \to q + \Lg$ contribution here, but the $\Lg \to q + \bar q$ contribution vanishes in the soft limit.) 

We note that the fact that the soft region does not contribute follows from the inclusive nature of our observable. With other observables, there would be a leading soft gluon contribution.

We note also that there is an angular ordering cancellation contained the soft limit splitting kernel, Eq.~(\ref{eq:softlimit}). If we did {\em not} have the factor (\ref{eq:subtractedfactor}), the important integration region for $\xi$ when $\tan^2(\vartheta/2) \ll 1$ would be $\xi\tan^2(\vartheta/2) \gtrsim 1$. That is, $\tan^2(\vartheta_{m+1}/2) \lesssim \tan^2(\vartheta/2)$. However, this effective angular ordering limit on the $\xi$ integration does not play a role in the present analysis.

This analysis has covered a virtuality ordered shower. For a transverse momentum ordered shower, the variable $y_T = z(1-z)y$ that was introduced in Sec.~\ref{sec:PTordered} is fixed by the evolution time. The wide-angle soft  limit is $y_T \to 0$ at fixed $\xi'$ where $(1-z) = \sqrt{\xi' y_T}$. Then the analysis just given for virtuality ordering applies also to transverse momentum ordering with $\sqrt{y_T}$ playing the role of $y$.

\section{Catani-Seymour shower scheme}
\label{CataniSeymour}

We now examine what happens with a Catani-Seymour dipole shower as defined in Ref.~\cite{Ringberg}. A parton with momentum $p_l$ splits to daughter partons with momenta $\hat p_l$ and $\hat p_{m+1}$. There is a helper parton with momentum $p_k$ that absorbs some momentum from the splitting, so that after the splitting its momentum is $\hat p_k$. The splitting is described by splitting variables $y$, $z$, and $\phi$. In contrast to the momentum mapping of Ref.~\cite{NSI}, in the Catani-Seymour scheme, the momenta of all other partons are unchanged by the splitting, $\hat p_i = p_i$ for $i \notin \{l,k\}$. The momenta $\hat p_l$, $\hat p_{m+1}$, and $\hat p_k$ are given in terms of $p_l$, and $p_k$ and the splitting variables by
\begin{equation}
\begin{split}
\label{eq:CSmomenta}
\hat p_l ={}& 
z\, p_l + y(1-z)\, p_k
+[2y z (1-z)\, p_l\cdot p_k]^{1/2}\kappa_\perp
\;\;,
\\
\hat p_{m+1} ={}& 
(1-z)\, p_l + yz\, p_k
-[2y z (1-z)\, p_l\cdot p_k]^{1/2}\kappa_\perp
\;\;,
\\
\hat p_k ={}& 
(1-y)\, p_k
\;\;.
\end{split}
\end{equation}
Here $\kappa_\perp$ is a spacelike unit vector orthogonal to $p_l$ and $p_k$ that has azimuthal angle $\phi$. Note that $y$ is proportional to the virtuality of the splitting
\begin{equation}
y = \frac{\hat p_l\cdot \hat p_{m+1}}{p_l\cdot p_k}
\;\;.
\end{equation}

For the observables that we need to analyze, we need some additional variables involving $Q$,
\begin{equation}
\begin{split}
\label{eq:CSkinematics}
\hat x ={}& \frac{2 \hat p_l\cdot Q}{Q^2}
\;\;,
\\
x ={}& \frac{2 p_l\cdot Q}{Q^2}
\;\;,
\\
1 - \cos\vartheta ={}& \frac{p_l\cdot p_k\, Q^2}{p_l\cdot Q\ p_k\cdot Q}
\;\;,
\\
r ={}& \frac{p_k\cdot Q}{p_l\cdot Q}
\;\;.
\end{split}
\end{equation}
Using the first equation in Eq.~(\ref{eq:CSmomenta}), we find how $\hat x$ is determined by $x$, $r$, $\vartheta$ and the splitting variables,
\begin{equation}
\hat x = Z(z,\phi,y;r,\vartheta)\, x
\;\;,
\end{equation}
where
\begin{equation}
Z(z,\phi,y;r,\vartheta) =
z + y (1-z) r 
- \sqrt{y z (1-z) r}\,\sin(\vartheta/2) \cos\phi
\;\;.
\end{equation}
Note that $Z(z,\phi,y,r,\vartheta) \to z$ for $y \to 0$. 

In defining the shower evolution, we take a spin average everywhere, so that spin indices disappear. In addition, we use the leading color approximation. Then we have just one color label for the $m$ parton state, $\{c\}_m$. The evolution equation is
\begin{equation}
\begin{split}
\label{eq:CSevolution1}
\frac{d}{dt}\sbrax{\hat x,\hat f_\La}\sket{\rho(t)}
={}&
\sum_m \frac{1}{m!}\int \big[d\{p,f,c\}_{m}\big]
\\ &\times
\sbra{\hat x,\hat f_\La}
{\cal H}^{\rm CS}_\LI(t) - {\cal V}^{\rm CS}(t)
\sket{\{p,f,c\}_{m}}
\\ & \times
\sbrax{\{p,f,c\}_{m}}
\sket{\rho(t)}
\;\;,
\end{split}
\end{equation}
where we integrate over a complete set of parton states $\sket{\{p,f,c\}_{m}}$ just before the splitting represented by the real and virtual splitting operators ${\cal H}^{\rm CS}_\LI(t)$ and ${\cal V}^{\rm CS}(t)$.

The matrix element of the splitting operator for a Catani-Seymour shower has the form
\begin{equation}
\begin{split}
\label{eq:HVxfCS}
\sbra{\hat x,\hat f_\La}&
{\cal H}_\LI^{\rm CS}(t)
-{\cal V}^{\rm CS}(t)
\sket{\{p,f,c\}_{m}}
=
\\&
\sum_l\sum_{k \ne l}
\sum_{\zeta_{\rm f}\in \Phi_{l}(f_l)}
\frac{2 p_l\cdot p_k}{16 \pi^2}
\int_{0}^{1}dy\, (1-y)\int_{0}^{1} dz\int_{0}^{2\pi}\frac{d\phi}{2\pi}
\\&\times
\delta\!\left(
t - T_l(\{\hat p,\hat f\}_{m+1})
\right)
\\&\times 
\bigg[
\sum_{i=1}^{m+1}
\delta_{\hat f_\La,\hat f_i}\,
\delta\!\left(\hat x - \frac{2\hat p_i\cdot Q}{Q^2}\right)
-
\sum_{i=1}^m
\delta_{\hat f_\La,f_i}\,
\delta\!\left(\hat x - \frac{2 p_i\cdot Q}{Q^2}\right)
\bigg]
\\&\times
\chi_{lk}(\{c\}_{m})\
w_{lk}^{\rm CS}(\{\hat f,\hat p\}_{m+1})
\;\;.
\end{split}
\end{equation}
For each splitting of a parton $l$ we choose a partner parton $k$. The momentum mapping depends on both $l$ and $k$, so the sum over $k$ comes to the outside. There is a sum over flavor choices $\zeta_\Lf$ for the splitting. If $f_l$ is a quark or antiquark flavor, there is only one choice, $f_l \to f_l + \Lg$. When $f_l = \Lg$ then the gluon can split to two gluons or to any quark-antiquark pair. There is an integration over the splitting variables $y$, $z$, $\phi$ with the jacobian that follows from the Catani-Seymour kinematics. There is a delta function that defines the shower time $t$. We will make the simplest choice of a virtuality ordered shower,
\begin{equation}
T_l(\{\hat p,\hat f\}_{m+1}) = 
\log\left(\frac{Q^2}{2 \hat p_l\cdot \hat p_{m+1}}\right)
= \log\left(\frac{Q^2}{y\ 2 p_l\cdot p_k}\right)
\;\;.
\end{equation}
Thus the $y$-integration can be eliminated, with
\begin{equation}
y = \frac{Q^{2}_{0}}{2p_{l}\cdot p_{k}} e^{-t}\;\;. 
\end{equation}
In the next factor in Eq.~(\ref{eq:HVxfCS}), the first term is from the splitting operator, ${\cal H}_\LI^{\rm CS}(t)$, while the second term is from the virtual splitting operator, ${\cal V}^{\rm CS}(t)$. Both operators contain the same splitting function, represented in the last line.

In all cases except a $\Lg \to q + \bar q$ splitting, the splitting function includes the possibility of emitting a gluon with label $m+1$ from a parton with label $i$ and emitting the gluon from a parton with label $j$, treated coherently. The total emission probability is divided into a term associated with parton $i$, for which we use $l=i$ and $k = j$ and another contribution associated with parton $j$ for which we use $l=j$ and $k = i$. The splitting function is $w_{lk}^{\rm CS}$. For $q \to q + \Lg$,
\begin{equation}
w_{lk}^{\rm CS}=
\frac{4\pi\alpha_\Ls}{y\,p_l\cdot p_k}\,
C_\LF \left[\frac{2}{1-z + zy} - (1+z)\right]
\;\;.
\end{equation}
Finally, $\chi_{lk}(\{c\}_{m}) = 1$ if partons $l$ and $k$ are color connected in color state $\{c\}_{m}$. Otherwise $\chi_{lk}(\{c\}_{m}) = 0$.

We now specialize to the case that $\hat f_\La$ is a quark flavor. The case of a $\Lg \to q + \bar q$ splitting is not particularly instructive, so we consider only a $q \to q + \Lg$ splitting. Then
\begin{equation}
\begin{split}
\label{eq:CSevolutionqqg1}
\left[
\frac{d}{dt}\sbrax{\hat x,\hat f_\La}\sket{\rho(t)}
\right]_{q \to q + \Lg}
={}&
\int_{0}^{1} dz\int_{0}^{2\pi}\frac{d\phi}{2\pi}\
\frac{\alpha_\Ls}{2\pi}\,
C_\LF \,(1-y)\left[\frac{2}{1-z + zy} - (1+z)\right]
\\ & \times
\int_{-1}^1\!d\cos\vartheta
\int_0^\infty\!dr
\\&\times 
\left[
\frac{\sbrax{\hat x/Z(z,\phi,y;r,\vartheta),\hat f_\La,r,\vartheta}\sket{\rho(t)}}
{Z(z,\phi,y;r,\vartheta)}\,
-
\sbrax{\hat x,\hat f_\La,r,\vartheta}
\sket{\rho(t)}
\right]
\;\;,
\end{split}
\end{equation}
where
\begin{equation}
\begin{split}
\label{eq:correlationCS}
\sbrax{x,\hat f_\La,r,\vartheta}\sket{\rho(t)}
={}&
\sum_m \frac{1}{m!}\int \big[d\{p,f,c\}_{m}\big]
\sum_l\sum_{k \ne l}\delta_{\hat f_\La,f_l}\,\chi_{lk}(\{c\}_{m})
\\ & \times 
\delta\!\left(x - \frac{2 p_l\cdot Q}{Q^2}\right)\,
\delta\!\left(
\cos\vartheta
-1
+\frac{Q^2\, p_l\cdot p_k}{Q\cdot p_l\ Q\cdot p_k}
\right)
\delta\!\left(r - \frac{p_k\cdot Q}{p_l\cdot Q}\right)\,
\\ & \times 
\sbrax{\{p,f,c\}_{m}}
\sket{\rho(t)}
\;\;.
\end{split}
\end{equation}
Here $\sbrax{x,\hat f_\La,r,\vartheta}\sket{\rho(t)}$ is a two parton correlation function. One parton, with flavor $\hat f_\La$, carries energy fraction $x$. Another parton, color connected to the first, is separated from it by an angle $\vartheta$ and carries energy fraction $r\,x$.

We can now make the approximation of strongly ordered virtualities in the shower. For this, we simply take the $y \to 0$ limit. In this limit, the splitting function simplifies and $Z(z,\phi,y;r,\vartheta) \sim z$. Then
\begin{equation}
\begin{split}
\label{eq:CSevolutionqqg2}
\frac{d}{dt}\sbrax{\hat x,\hat f_\La}\sket{\rho(t)}
\approx{}&
\int_{0}^{1} dz\int_{0}^{2\pi}\frac{d\phi}{2\pi}\
\frac{\alpha_\Ls}{2\pi}\,
C_\LF \left[\frac{2}{1-z} - (1+z)\right]
\int_{-1}^1\!d\cos\vartheta
\int_0^\infty\!dr
\\&\times 
\left[
\frac{1}{z}
\sbrax{\hat x/z,\hat f_\La,r,\vartheta}\sket{\rho(t)}
-
\sbrax{\hat x,\hat f_\La,r,\vartheta}
\sket{\rho(t)}
\right]
\;\;.
\end{split}
\end{equation}
Now nothing depends on $\phi$, so we can trivially perform the $\phi$-integration. Furthermore, the integral over $\cos\vartheta$ and $r$ of the two parton correlation function is the one parton distribution function,
\begin{equation}
\int_{-1}^1\!d\cos\vartheta
\int_0^\infty\!dr\
\sbrax{x,\hat f_\La,r,\vartheta}
\sket{\rho(t)}
= \sbrax{x,\hat f_\La}
\sket{\rho(t)}
\;\;.
\end{equation}
Thus we obtain, within the strongly ordered virtualities approximation, the $q \to q + \Lg$ contribution to the DGLAP equation for the energy distribution function of a quark,
\begin{equation}
\begin{split}
\label{eq:CSevolutionqqg3}
\left[
\frac{d}{dt}\sbrax{\hat x,\hat f_\La}\sket{\rho(t)}
\right]_{q \to q + \Lg}
\approx{}&
\frac{\alpha_\Ls}{2\pi}
\int_{0}^{1} dz\
C_\LF\ \frac{1+z^2}{1-z}
\left[
\frac{1}{z}
\sbrax{\hat x/z,\hat f_\La}\sket{\rho(t)}
-
\sbrax{\hat x,\hat f_\La}
\sket{\rho(t)}
\right]
\;\;.
\end{split}
\end{equation}
This is often written using the notation
\begin{equation}
\begin{split}
\label{eq:CSevolutionqqg4}
\left[
\frac{d}{dt}\sbrax{\hat x,\hat f_\La}\sket{\rho(t)}
\right]_{q \to q + \Lg}
\approx{}&
\frac{\alpha_\Ls}{2\pi}
\int_{0}^{1} \frac{dz}{z}\,
C_\LF \left[\frac{1+z^2}{1-z}\right]_+\
\sbrax{\hat x/z,\hat f_\La}\sket{\rho(t)}
\;\;.
\end{split}
\end{equation}

We have presented the derivation for a virtuality ordered shower. The Catani-Seymour dipole showers of Refs.~\cite{Weinzierl,Schumann,SjostrandSkands} use transverse momentum as the ordering variable,
\begin{equation}
\label{eq:TdefPTbis}
T_l(\{\hat p,\hat f\}_{m+1}) = \log\left(
\frac{Q^2}{2z(1-z)\hat p_l\cdot \hat p_{m+1}}
\right)
\;\;.
\end{equation}
Then the variable
\begin{equation}
y_T = z (1-z) y
= \frac{Q^2}{2 p_l\cdot p_k}\ e^{-t}
\end{equation}
is fixed by the delta function $\delta(t - T_l)$. The strong ordering limit is then $y_T \to 0$ at fixed $z$. In this case, since the definition of $y$ entails that $y < 1$, one should include a factor
\begin{equation}
\theta(z(1-z) > y_T)
\end{equation}
in the integrand. However, this factor disappears in the limit $y_T \to 0$, leaving us with the same result (\ref{eq:CSevolutionqqg3}).

In our derivation, we have simply taken the limit $y \to 0$ (or $y_T \to 0$) inside of the integration over $z$. One could explicitly examine the region in the $z$-integration that corresponds to soft gluon emission, $y\to 0$ with $(1-z)/y \equiv \xi$ finite. To do that, one changes variables from $z$ to $\xi$ and takes the $y\to 0$ limit inside the integration over $\xi$. This leads to what would be a finite contribution to the integral except for the factor
\begin{equation}
\left[
\frac{\sbrax{\hat x/Z(z,\phi,y;r,\vartheta),\hat f_\La,r,\vartheta}\sket{\rho(t)}}
{Z(z,\phi,y;r,\vartheta)}\,
-
\sbrax{\hat x,\hat f_\La,r,\vartheta}
\sket{\rho(t)}
\right]
\;\;.
\end{equation}
For a virtuality ordered shower, this factor provides a factor $y$ that removes the soft emission contribution in the $y\to 0$ limit at fixed $\xi = (1-z)/y$. For a transverse momentum ordered shower, this factor provides a factor $\sqrt{y_T}$ that removes the soft emission contribution in the $y_T\to 0$ limit at fixed $\xi' \equiv \xi/z =(1-z)^2/y_T$. This justifies taking the limit $y \to 0$ (or $y_T \to 0$) inside of the original integration over $z$.

\section{Conclusions}
\label{sec:conclusions}

In a final state parton shower, the distribution of the energy carried by a single parton evolves as the shower evolves and the energy carried by each mother parton is distributed among the daughters. We have examined this evolution for a final state parton shower in electron-positron annihilation. Our principle example is based on the shower evolution equations of Ref.~\cite{NSI}, which include complete color and spin information. These evolution equations are based on a dipole picture of parton splitting, in which gluon emission from a parton $l$ can interfere with gluon emission from another parton $k$. We have also analyzed evolution in (spin averaged, leading color) shower evolution based on the Catani-Seymour splitting functions and momentum mapping, as used in the practical shower programs of Refs.~\cite{Weinzierl,Schumann} and in the final state showers of \textsc{Pythia} \cite{SjostrandSkands, Pythia8}.

The use of a parton shower to model the evolution of the final state created from a short distance scattering process is based on the idea that successive steps in the evolution are separated by happening on very different time scales. That is, each successive step involves much smaller virtuality (or transverse momentum) than the previous step. The shower evolution is based on factoring the amplitude for a relatively soft splitting from the previous harder splittings. The actual evolution defined by the evolution equation includes integrations over the virtuality of each splitting, including an integration region in which the virtuality is {\em not} much smaller than the previous virtuality.\footnote{For this reason, one attempts to define the splitting functions and momentum mapping in such a way that the parton shower model does something sensible when the ratio of successive splitting virtualities is not small.} Whatever is left over coming from the region in which the virtualities of successive splittings of the same parton are similar rather than strongly ordered may be considered to contribute to the second order DGLAP kernel. Thus this contribution is suppressed by a factor of $\alpha_\Ls$.

We have examined the evolution of the distribution of the energy carried by a parton, $\sbrax{x,f_\La}\sket{\rho(t)}$, as defined in Eq.~(\ref{eq:xfrhodef}). The analysis presented in Sec.~\ref{sec:evolution} for the shower evolution equations of Ref.~\cite{NSI} covered the evolution of the energy distribution for quarks. For the sake of completeness, we include the evolution of the energy distribution for gluons in Appendix \ref{sec:Eingluons}. For evolution using  the Catani-Seymour splitting functions and momentum mapping, we have limited the analysis to the evolution of the energy distribution for quarks from $q \to q + \Lg$ splitting. In each case, we emphasized a virtuality ordered shower, but pointed out that a transverse momentum ordered shower gives the same result for the evolution of the parton energy distribution. In each case examined, we find that the energy distribution obeys the standard lowest order DGLAP equation when we approximate the evolution using the strong ordering limit. This result addresses the concern expressed by Dokshitzer and Marchesini \cite{Dokshitzer} that evolution in dipole based showers might fail to follow the DGLAP equation.

The shower evolution time $t$ represents the scale $\mu^2 = 2 \hat p_l\cdot \hat p_{m+1}$ at which one parton is resolved into two daughter partons. When $\mu^2$ is much smaller than the scale $Q^2$ of the hard process, the logarithm $t = \log(Q^2/\mu^2)$ is large. When $\sbrax{x,f_\La}\sket{\rho(t)}$ is expanded in powers of $\alpha_\Ls$, the perturbative coefficients contain powers of $t$. The evolution equation enables us to sum the large logarithms of $Q^2/\mu^2$. There are a number of other cases in which a physical observable presents us with a large ratio of physical scales and in which it is known how to sum the large logarithms of this ratio. In some of these cases, parton shower evolution will generate the correct summation. This is what we found here for the parton energy distribution, which, although not directly observable, maps to the observable hadron energy distribution. In some other cases, it is pretty straightforward to see that a parton shower generates the correct summation, as for the dependence of jet cross sections on a jet resolution parameter based on $k_T$ when the jets are generated with a $k_T$ ordered parton shower. In other cases, parton shower evolution will generate a summation of large logarithms, but it may be the wrong summation. For instance, we expect that standard parton showers do not generate ``BFKL'' logarithms that describe the physics of large rapidity differences, as in \cite{Jeppe}. 

For a given choice of shower evolution it is of significant interest to know what summations of large logarithms the shower evolution equation correctly produces. That is, one would like to validate particular shower evolution schemes against known results for summations of large logarithms. Consider, for example the energy-energy correlation function in electron-positron annihilation at nearly back-to-back angles $\theta$. One knows the summation of large logarithms $\log(1/\theta^2)$ in full QCD. Then one needs to derive the corresponding summation in the shower model, deriving the appropriate evolution equation from the shower evolution equation. A second example in electron-positron annihilation is the summation of logarithms of $1-x$ in the energy distribution of partons or hadrons. For hadron-hadron collisions, one should investigate, for example, the transverse momentum distribution of Drell-Yan lepton pairs. 

We note that the existing literature goes a long way toward validating parton shower evolution against known large logarithm summations in the case of parton showers based on using the angle between daughter partons in a splitting as the shower evolution variable. Indeed, some of the derivations of large logarithm summations in QCD make use of on an angle-ordered parton cascade picture \cite{ErmolaevQCDSummation, MuellerQCDSummation, CataniQCDSummation, Catani32etalQCDsummation, CacciariQCDSummation}. Correspondingly, the paper \cite{CataniMCsummation} argues that a parton shower ordered in splitting angles gets the leading terms in certain large logarithm summations correctly. 

To our knowledge, less is known analytically about how well virtuality or $k_T$ ordered showers reproduce known large logarithm summations. We do note that Ref.~\cite{BanfiMCsummation} looks at large logarithm summations for both angle ordered and for virtuality ordered or $k_T$ ordered showers with respect to ``non-global'' observables. 

To build from the existing results, one would need to check carefully, for a given shower scheme, the accuracy (in powers of $\as$ and the logarithms) with which the full QCD result for a given kind of summation is reproduced by the shower. As emphasized by Dokshitzer and Marchesini \cite{Dokshitzer}, one would need to pay close attention to the treatment of momentum conservation in the shower. 

We do not attempt a general analysis here. What we offer in this paper applies only to the circumscribed problem suggested in Ref.~\cite{Dokshitzer}. For further investigations along these lines, we believe that it is useful to have shower evolution expressed as a definite integral equation, as in Ref.~\cite{NSI}. We hope that the present paper provides some hints about how a program of validating parton shower evolution against known large logarithm summations could be pursued.


\acknowledgments{
We are grateful to Y.~Dokshitzer and G.~Marchesini for helpful conversations and correspondence. Part of this work was carried out while the authors were at the theory group at CERN, along with Dokshitzer and Marchesini. We are grateful to CERN for its support. We are also grateful to  P.~Skands, S.~Weinzierl, and G.~Gustafson for helpful correspondence. 
This work was supported in part by the United States Department of Energy, the Helmoltz Alliance network ``Physics at the Terascale," and the Hungarian Scientific Research Fund grant OTKA K-60432.
}

\appendix

\section{Energy distribution of gluons}
\label{sec:Eingluons}

We have analyzed the evolution of the energy distribution $\sbrax{x,f_\La}\sket{\rho(t)}$ in the case that $f_\La$ is a quark flavor. The case of an antiquark flavor is so similar that it hardly needs a separate treatment. However, there are some differences in the case that $f_\La = \Lg$. We provide a brief analysis of this case in this appendix.

There are three sorts of contributions. The first is a $q \to q + \Lg$ or $\bar q \to \bar q + \Lg$ splitting, as depicted in Fig.~\ref{fig:qqgdirect}, in which it is the energy of the daughter gluon, instead of the daughter quark or antiquark, that is measured. The second is a $\Lg \to \Lg + \Lg$ splitting, as in Fig.~\ref{fig:gqq} but with the $q+ \bar q$ replaced by $\Lg + \Lg$. Either of the two daughter gluons can be the one whose energy is measured. There is a virtual splitting counter term for this contribution. There are $\Lg \to \Lg + \Lg$ interference graphs, but these are singular only in the soft limit and do not give leading contributions for the evolution of the parton energy distributions, as we have seen in some detail for quark splitting. (Evidently, if it is the soft gluon itself that is measured then there is no contribution to the evolution of $\sbrax{x,f_\La}\sket{\rho(t)}$ at any finite $x$ in the limit in which the soft gluon energy vanishes.) The final contribution is the virtual splitting counter term for a $\Lg \to q + \bar q$ splitting, as depicted in Fig.~\ref{fig:gqq}.

For the gluon case, Eq.~(\ref{eq:HVxf5}) becomes
\begin{equation}
\begin{split}
\label{eq:gluon1}
\sbra{\hat x,\Lg}&
{\cal H}_\LI(t)
-{\cal V}(t)
\sket{\{p,f,s',c',s,c\}_{m}}
\approx
\\&
\sum_l
\frac{p_l\cdot Q}{8\pi^2}\ y
\int_0^1\!dz\, \int \!\frac{d\phi}{2\pi}
\int_0^1\! dx \ 
\delta\!\left(x - \frac{2 p_l\cdot Q}{Q^2}\right)\,
\sum_{\hat f_l,\hat f_{m+1}} 
\\&\times
\brax{\{s'\}_{m}}\ket{\{s\}_{m}}\
\brax{\{c'\}_{m}}\ket{\{c\}_{m}}
\\&\times
\biggl\{
\theta(f_{l} \ne {\rm g},\hat f_{l} = f_l, \hat f_{m+1} = \Lg)\
C_\LF\,
\overline w_{ll}(\{\hat f,\hat p\}_{m+1})\,
\frac{1}{1-z}\,
\delta\!\left(x - \frac{\hat x}{1-z}\right)
\\&\quad
+
\theta(f_{l} = {\rm g},\hat f_{l} = \Lg, \hat f_{m+1} = \Lg)\
C_\LA\,\overline w_{ll}(\{\hat f,\hat p\}_{m+1})
\\ &\qquad\times
\bigg[
\frac{1}{1-z}\,
\delta\!\left(x - \frac{\hat x}{1-z}\right)
+\frac{1}{z}\,
\delta\!\left(x - \frac{\hat x}{z}\right)
-
\delta\!\left(x - \hat x\right)
\bigg]
\\&\quad
-
\theta(f_{l} = {\rm g},\hat f_{l} \in {\cal Q}, \hat f_{m+1} = -\hat f_l)\
T_\LR\,\overline w_{ll}(\{\hat f,\hat p\}_{m+1})\
\delta\!\left(x - \hat x\right)
\bigg\}
\;\;.
\end{split}
\end{equation}
There are five terms in three groups. In each case, we have adopted the notation that parton $m+1$ carries momentum fraction $1-z$. In the first term, the mother parton with flavor $f_l$ is a quark or antiquark that splits into a parton of the same flavor with index $l$ and a gluon with index $m+1$. We measure the energy of the daughter gluon, so $\hat x = (1-z) x$. In the next three terms, the mother parton $l$ is a gluon and daughter partons $l$ and $m+1$ are both gluons. The definitions of Ref.~\cite{NSI} explicitly break the symmetry between gluons $l$ and $m+1$ in such a way that there is a singularity in $\overline w_{ll}$ corresponding to gluon $m+1$ being soft, but no singularity corresponding to gluon $l$ being soft. In the second term, we measure the energy of the daughter with label $m+1$, so $\hat x = (1-z) x$. In the third term, we measure the energy of the daughter with label $l$, so $\hat x = z x$. The fourth term is the virtual splitting counter term, with $\hat x = x$. The fifth term is the virtual splitting counter term for a $\Lg \to q + \bar q$ splitting. Here $\hat f_{l} \in {\cal Q}$ means that $\hat f_{l}$ is a quark flavor, ${\cal Q} = \{{\rm u}, {\rm d}, {\rm s}, \dots\}$. 

All that we need is the splitting functions $\overline w_{ll}$ in the $y \to 0$ limit. For $f_l \ne g$ we have
\begin{equation}
\overline w_{ll} \sim 
\frac{4\pi \alpha_\Ls}{p_l\cdot Q\ y}\
\frac{1 + z^2}{1-z}
\;\;,
\end{equation}
as in Eqs.~(\ref{eq:wll}) and (\ref{eq:splittingFlimits}). For $f_l = \Lg$, $\hat f_l = \Lg$, we have, from Eqs.~(2.46) and (2.52) of Ref.~\cite{NSII}\footnote{We interchange $z$ and $(1-z)$ to match the conventions of this paper.},
\begin{equation}
\overline w_{ll} \sim 
\frac{4\pi \alpha_\Ls}{p_l\cdot Q\ y}\
\left[\frac{2z}{1-z} + z(1-z)\right]
\;\;.
\end{equation}
For $f_l = \Lg$, $\hat f_l \in {\cal Q}$, we have
\begin{equation}
\overline w_{ll} \sim 
\frac{4\pi \alpha_\Ls}{p_l\cdot Q\ y}\
[z^2 + (1-z)^2]
\;\;,
\end{equation}
as in Eqs.~(\ref{eq:wll}) and (\ref{eq:splittingFlimits}). We make these substitutions and perform the trivial $\phi$ integration. In the first two of the five terms, we use the freedom to change integration variables to interchange the roles of $z$ and $1-z$. This gives
\begin{equation}
\begin{split}
\label{eq:gluon2}
\sbra{\hat x,\Lg}&
{\cal H}_\LI(t)
-{\cal V}(t)
\sket{\{p,f,s',c',s,c\}_{m}}
\approx
\\&
\sum_l
\int_0^1\!dz
\int_0^1\! dx \ 
\delta\!\left(x - \frac{2 p_l\cdot Q}{Q^2}\right)
\brax{\{s'\}_{m}}\ket{\{s\}_{m}}\
\brax{\{c'\}_{m}}\ket{\{c\}_{m}}
\\&\times
\biggl\{
\theta(f_{l} \ne {\rm g})\
\frac{\alpha_\Ls}{2\pi}\
C_\LF\,
\frac{1 + (1-z)^2}{z}\,
\frac{1}{z}\,
\delta\!\left(x - \frac{\hat x}{z}\right)
\\&\quad
+
\theta(f_{l} = \Lg)\
\frac{\alpha_\Ls}{2\pi}\
C_\LA\,
\left[\frac{2(1-z)}{z} + z(1-z)\right]\,
\frac{1}{z}\,
\delta\!\left(x - \frac{\hat x}{z}\right)
\\&\quad
+
\theta(f_{l} = \Lg)\
\frac{\alpha_\Ls}{2\pi}\
C_\LA\,
\left[\frac{2z}{1-z} + z(1-z)\right]
\bigg[
\frac{1}{z}\,
\delta\!\left(x - \frac{\hat x}{z}\right)
-
\delta\!\left(x - \hat x\right)
\bigg]
\\&\quad
-
\theta(f_{l} = \Lg)\
\frac{\alpha_\Ls}{2\pi}\
T_\LR\,n_\Lf\,[z^2 + (1-z)^2]\,
\delta\!\left(x - \hat x\right)
\bigg\}
\;\;.
\end{split}
\end{equation}
In the last line, $n_\Lf$ is the number of quark flavors, that is, the number of elements of ${\cal Q}$.

We can now substitute this result into Eq.~(\ref{eq:xfevolution1}) for the evolution of $\sbrax{\hat x,\Lg}\sket{\rho(t)}$. On the right hand side of the evolution equation, we use the definition (\ref{eq:xfrhodef}) of $\sbrax{\hat x,\hat f_\La}\sket{\rho(t)}$. This gives
\begin{equation}
\begin{split}
\label{eq:xfevolutionglue}
\frac{d}{dt}\sbrax{\hat x,\Lg}\sket{\rho(t)}
={}&
\frac{\alpha_\Ls}{2\pi}
\int_0^1\!dz\
\biggl\{
\sum_{f' \ne {\rm g}}\
C_\LF\,
\frac{1 + (1-z)^2}{z}\,
\frac{1}{z}\,
\sbrax{\hat x/z,f'}\sket{\rho(t)}
\\&\quad
+
C_\LA\,
\left[\frac{2(1-z)}{z} + z(1-z)\right]\,
\frac{1}{z}\,
\sbrax{\hat x/z,\Lg}\sket{\rho(t)}
\\&\quad
+
C_\LA\,
\left[\frac{2z}{1-z} + z(1-z)\right]
\bigg[
\frac{1}{z}\,
\sbrax{\hat x/z,\Lg}\sket{\rho(t)}
-
\sbrax{\hat x,\Lg}\sket{\rho(t)}
\bigg]
\\&\quad
- T_\LR\,n_\Lf\,[z^2 + (1-z)^2]\,
\sbrax{\hat x,\Lg}\sket{\rho(t)}
\bigg\}
\;\;.
\end{split}
\end{equation}
This is the DGLAP equation. It is usually written as
\begin{equation}
\begin{split}
\label{eq:xfevolutionglue2}
\frac{d}{dt}\sbrax{\hat x,\Lg}\sket{\rho(t)}
={}&
\frac{\alpha_\Ls}{2\pi}
\int_{\hat x}^1\!\frac{dz}{z}\
\biggl\{
\sum_{f' \ne {\rm g}}\
C_\LF\,
\frac{1 + (1-z)^2}{z}\,
\sbrax{\hat x/z,f'}\sket{\rho(t)}
\\&\quad
+
\bigg[2 C_\LA \left(\frac{z}{[1-z]_+} + \frac{(1-z)}{z} + z(1-z)\right)
\\ & \qquad\quad
+ \frac{11 C_\LA - 4T_\LR\,n_\Lf}{6}\ \delta(1-z)
\bigg]
\sbrax{\hat x/z,\Lg}\sket{\rho(t)}
\bigg\}
\;\;.
\end{split}
\end{equation}
%


\end{document}